\documentclass[runningheads]{llncs} %

\usepackage[utf8]{inputenc}
\usepackage{amsmath,amssymb}
\usepackage[bookmarks=true]{hyperref} %
\usepackage{cleveref}
\usepackage{tabularx}
\usepackage{booktabs} %
\usepackage{multirow} 
\usepackage{paralist}
\usepackage{wrapfig}

\hypersetup{
pdftitle ={Maximizing Reachability  Probabilities in Rectangular Automata with Random Clocks},
pdfauthor ={Joanna Delicaris, Stefan Schupp, Erika \'{A}brah\'{a}m, 
Anne Remke},
bookmarksopen=true
}

\usepackage{xifthen}

\usepackage{soul}
\newif\ifistoreview
\newif\ifhighlightchanges
\newif\ifrespectstatus
\istoreviewfalse %
\highlightchangesfalse %
\respectstatusfalse %

\newcommand{\beginreview}{\ifrespectstatus\highlightchangesfalse\istoreviewfalse\fi}
\newcommand{\reviewstatus}{\ifrespectstatus\highlightchangestrue\fi}

\ifrespectstatus
\istoreviewfalse
\highlightchangesfalse
\fi

\usepackage[]{todonotes}

\usepackage{tikz} 
\usetikzlibrary{arrows,automata,backgrounds,positioning,intersections,shapes,decorations.pathreplacing,positioning, patterns}
\usepackage[nointegrals]{wasysym}
\usepackage{tkz-euclide}
\usepackage{subcaption}
\usepackage{graphicx}
\usepackage{xspace}
\usepackage{pgfplots}
\pgfplotsset{compat=newest}

\usepackage{algorithmic}
\usepackage{algorithm}
\usepackage{listings}
\algsetup{linenosize=\footnotesize}

\usepackage{mathtools}
\spnewtheorem*{runningexample}{Running example}{\itshape}{\rmfamily}
\usepackage[binary-units]{siunitx}
\usepackage{nicefrac}

\usepackage[ligature, inference]{semantic}

\usepackage{xcolor}
\definecolor{darkgreen}{rgb}{0.0, 0.42, 0.24}
\definecolor{solBlue}{RGB}{38 139 210}
\definecolor{amethyst}{rgb}{0.6, 0.4, 0.8}
\definecolor{antiquefuchsia}{rgb}{0.57, 0.36, 0.51}
\definecolor{darkorchid}{rgb}{0.6, 0.2, 0.8}
\definecolor{darkviolet}{rgb}{0.58, 0.0, 0.83}
\definecolor{deepmagenta}{rgb}{0.8, 0.0, 0.8}
\definecolor{deeplilac}{rgb}{0.6, 0.33, 0.73}
\definecolor{electricpurple}{rgb}{0.75, 0.0, 1.0}
\definecolor{electricviolet}{rgb}{0.56, 0.0, 1.0}
\definecolor{mediumorchid}{rgb}{0.73, 0.33, 0.83}
\definecolor{patriarch}{rgb}{0.5, 0.0, 0.5}
\definecolor{phlox}{rgb}{0.87, 0.0, 1.0}
\definecolor{purple(munsell)}{rgb}{0.62, 0.0, 0.77}
\definecolor{purple(x11)}{rgb}{0.63, 0.36, 0.94}
\definecolor{alertcolor}{named}{purple(munsell)}

\makeatletter
\newcounter{groupcount}
\pgfplotsset{
    draw group line/.style n args={5}{
        after end axis/.append code={
            \setcounter{groupcount}{0}
            \pgfplotstableforeachcolumnelement{#1}\of\datatable\as\cell{%
                \def\temp{#2}
                \ifx\temp\cell
                    \ifnum\thegroupcount=0
                        \stepcounter{groupcount}
                        \pgfplotstablegetelem{\pgfplotstablerow}{[index]0}\of\datatable
                        \coordinate [yshift=#4] (startgroup) at (axis cs:\pgfplotsretval,0);
                    \else
                        \stepcounter{groupcount}
                        \pgfplotstablegetelem{\pgfplotstablerow}{[index]0}\of\datatable
                        \coordinate [yshift=#4] (endgroup) at (axis cs:\pgfplotsretval,0);
                    \fi
                \else
                    \ifnum\thegroupcount>1
                        \setcounter{groupcount}{0}
                        \draw [
                            shorten >=-#5,
                            shorten <=-#5
                        ] (startgroup) -- node [anchor=north] {#3} (endgroup);
                    \fi
                \fi
            }
            \ifnum\thegroupcount>1
                        \setcounter{groupcount}{0}
                        \draw [
                            shorten >=-#5,
                            shorten <=-#5
                        ] (startgroup) -- node [anchor=north] {#3} (endgroup);
            \fi
            \ifnum\thegroupcount=1
                        \setcounter{groupcount}{0}
                        \draw [
                            shorten >=-#5,
                            shorten <=-#5
                        ] ($(startgroup)-(1.5mm,0)$) -- node [anchor=north] {#3} ($(startgroup)+(1.5mm,0)$);
            \fi
        }
    }
}
\makeatother

\pgfplotsset{
tick label style={font=\footnotesize},
label style={font=\small},
legend style={font=\footnotesize},
}

\definecolor{solBase01}{RGB}{88 110 117}
\definecolor{solMagenta}{RGB}{211  54 130}
\definecolor{color2}{named}{solMagenta}
\definecolor{solGreen}{RGB}{133 153   0}
\definecolor{color3}{named}{solGreen}
\definecolor{solYellow}{RGB}{181 137   0}
\definecolor{color4}{named}{solYellow}
\definecolor{dartmouthgreen}{rgb}{0.05, 0.5, 0.06}
\definecolor{color5}{named}{dartmouthgreen}
\definecolor{solBlue}{RGB}{38 139 210}
\definecolor{color6}{named}{solBlue}
\definecolor{davysgrey}{rgb}{0.33, 0.33, 0.33}
\definecolor{fancyblue}{rgb}{0.01, 0.28, 1.0}
\definecolor{amethyst}{rgb}{0.6, 0.4, 0.8}

\definecolor{mainColor}{named}{solBase01}
\definecolor{boxcolor}{named}{color4}
\definecolor{refinementcolor}{named}{color2}
\definecolor{cinitloc}{named}{mainColor}
\definecolor{cpath1light}{named}{color6}
\definecolor{cpath1dark}{named}{fancyblue}
\definecolor{cpath2light}{named}{color3}
\definecolor{cpath2dark}{named}{color5}
\definecolor{crefseg}{named}{color2}
\definecolor{cinterseg}{named}{amethyst}

\tikzstyle{initloc}=[cinitloc]
\tikzstyle{path1start}=[cpath1light,densely dotted]
\tikzstyle{path1end}=[cpath1dark]
\tikzstyle{path2start}=[cpath2light,dotted]
\tikzstyle{path2end}=[cpath2dark]

\tikzstyle{filling}=[very thick, fill, fill opacity = 0.2]
\tikzstyle{cone}=[draw = none, fill, fill opacity = 0.2]
\tikzstyle{border}=[very thick, line join = round]
\tikzstyle{guard}=[fill opacity=0.1]
\tikzstyle{guardticks}=[color3]
\tikzstyle{init}=[border, mainColor, line join = round]
\tikzstyle{refinement}=[pattern=north west lines, pattern color = color2]
\tikzstyle{refinement2}=[pattern=north east lines, pattern color = color2]
\tikzstyle{other}=[opacity = 0.15]%

\tikzstyle{refinedsegment}=[very thick, line join = round, fill, fill opacity = 0.2, color2]
\tikzstyle{intermediatesegment}=[line join = round, fill, cinterseg]

\newcommand{\ie}{i.e.,\ }

\newcommand{\dimC}{\ensuremath{{d_{C}}}\xspace}
\newcommand{\dimR}{\ensuremath{{d_{R}}}\xspace}

\newcommand{\Reals}{\ensuremath{\mathbb{R}}\xspace}
\newcommand{\Realspos}{\ensuremath{\Reals_{>0}}\xspace}
\newcommand{\Realsneg}{\ensuremath{\Reals_{<0}}\xspace}
\newcommand{\Realsposzero}{\ensuremath{\Reals_{\geq0}}\xspace}

\newcommand{\Q}{\mathbb{Q}}

\newcommand{\Naturals}{\ensuremath{\mathbb{N}}\xspace}

\newcommand{\support}{\ensuremath{\mathit{supp}}\xspace}
\newcommand{\CDFs}{\ensuremath{\mathbb{F}_c}\xspace}
\newcommand{\DDFs}{\ensuremath{\mathbb{F}_d}\xspace}
\newcommand{\DFs}{\ensuremath{\mathbb{F}}\xspace}
\newcommand{\CDF}{\ensuremath{\mathit{distr}}\xspace}

\newcommand{\dur}{\mathit{dur}}

\newcommand{\pre}{\mathit{pre}}
\newcommand{\post}{\mathit{post}}
\newcommand{\reset}{\mathit{reset}}

\newcommand{\scheduler}{\ensuremath{\mathfrak{s}}\xspace}

\newcommand{\SchedulersProphetic}{\ensuremath{\Schedulers}\xspace} 
\newcommand{\Schedulers}{\ensuremath{\mathfrak{S}}\xspace}
\newcommand{\SchedulersProphMax}{\ensuremath{\Schedulers_{\textsl{goal}}}\xspace}
\newcommand{\SchedulersRepr}{{\ensuremath{[\scheduler]_i}}\xspace}
\newcommand{\SchedulersRepri}[1]{{\ensuremath{[\scheduler]_{#1}}}\xspace}

\newcommand{\myquad}{\hspace*{2ex}}
\newcommand{\AR}{\ensuremath{R}\xspace}
\newcommand{\A}{\ensuremath{\mathcal{A}}\xspace}
\newcommand{\Loc}{\ensuremath{\mathit{Loc}}\xspace} 
\newcommand{\VarAut}{\ensuremath{\mathit{Var}_{C}}\xspace}
\newcommand{\Var}{\VarAut}
\newcommand{\Edge}{\ensuremath{\mathit{Edge}_C}\xspace} 
\newcommand{\EdgeRandom}{\ensuremath{\mathit{Edge}_R}\xspace} 
\newcommand{\Act}{\ensuremath{\mathit{Flow}_C}\xspace}
\newcommand{\ActRandom}{\ensuremath{\mathit{Flow}_R}\xspace}
\newcommand{\Distr}{\ensuremath{\mathit{Distr}}\xspace}
\newcommand{\Inv}{\ensuremath{\mathit{Inv}}\xspace}
\newcommand{\Init}{\ensuremath{\mathit{Init}}\xspace} 

\newcommand{\VarRandom}{\ensuremath{\mathit{Var}_{R}}\xspace} 

\newcommand{\clocki}{\ensuremath{r}\xspace}
\newcommand{\sample}{\ensuremath{s}\xspace}
\newcommand{\randomvar}{\ensuremath{S}\xspace}
\newcommand{\randomvarij}{\ensuremath{S_{r,n}}\xspace}
\newcommand{\contrandomvar}{\ensuremath{r_{n}}\xspace}

\newcommand{\tmax}{\ensuremath{t_\textsl{max}}\xspace} %
\newcommand{\tint}{\ensuremath{t_\textsl{int}}\xspace} %
\newcommand{\jumpmax}{\ensuremath{\mathit{jmp}}\xspace} %

\newcommand{\States}{\ensuremath{\mathcal{S}}\xspace} %
\newcommand{\val}{\ensuremath{\nu}\xspace} %
\newcommand{\valRandom}{\ensuremath{\mu}\xspace}
\newcommand{\valSet}{\ensuremath{\mathcal{V}}\xspace} %
\newcommand{\Goal}{\ensuremath{\smash{\States^\textsl{goal}}}\xspace}

\newcommand{\Goallocations}{\ensuremath{L^\textsl{goal}}\xspace}

\newcommand{\Goalvalset}{\ensuremath{\valSet^\textsl{goal}}\xspace}
\newcommand{\Goali}{\ensuremath{\smash{\States^{i}}}\xspace}
\newcommand{\Goalvali}{\ensuremath{\smash{\valSet^{i}}}\xspace}
\newcommand{\Goalvalik}[1]{\ensuremath{\smash{\valSet^{i}_{#1}}}\xspace}
\newcommand{\Refvalik}[1]{\ensuremath{\smash{\hat{\valSet}^{i}_{#1}}}\xspace}

\newcommand{\pathindices}{\ensuremath{I_{\States}}\xspace}

\newcommand{\reachtree}{\ensuremath{\mathcal{R}}\xspace}

\newcommand{\estat}{\ensuremath{e_\textsl{stat}}\xspace}
\newcommand{\einfty}{\ensuremath{e_{\infty}}\xspace}
\newcommand{\comptime}{\ensuremath{t_\textsl{comp}}}
\newcommand{\compintervals}{\ensuremath{\#I}}

\newcommand{\RefinedSegmentik}[2]{\ensuremath{\smash{\hat{\valSet}^{#1}_{#2}}}\xspace}
\newcommand{\JumpSegmentik}[2]{\ensuremath{\smash{{\mathcal{Q}}^{#1}_{#2}}}\xspace}
\newcommand{\map}{\ensuremath{\smash{map_i}\xspace}}

\newcommand{\hastate}{\ensuremath{\sigma}\xspace} %
\newcommand{\Goalreached}{\ensuremath{\smash{\States^\textsl{goal}_\textsl{reach}}}\xspace}

\newcommand{\assignment}{\mathbf{s}} %

\newcommand{\Path}{\ensuremath{\pi}} %
\newcommand{\hahistory}{\Path} %
\newcommand{\Paths}{\ensuremath{\mathit{Paths}}\xspace} %
\newcommand{\hrep}{{\textit{H}}\textit{-re\-pre\-sen\-ta\-tion}\xspace} %
\newcommand{\vrep}{{\textit{V}}\textit{-re\-pre\-sen\-ta\-tion}\xspace} %

\newcommand{\EnabledJumps}{\ensuremath{\mathit{EnabledJumps}}\xspace}
\newcommand{\Time}{\ensuremath{\mathit{EnabledTime}}\xspace}

\newcommand{\EntryVal}[1][]{%
  \ifthenelse{\isempty{#1}}%
    {\ensuremath{E_{l}\xspace}}%
    {\ensuremath{E_{#1}\xspace}}%
}

\newcommand{\ftc}[1]{\ensuremath{\mathit{T^{+}_{#1}}}\xspace}
\newcommand{\btc}[1]{\ensuremath{\mathit{T^{-}_{#1}}}\xspace}
\newcommand{\fosr}[1]{\ensuremath{\mathit{D^{+}_{#1}}}\xspace}
\newcommand{\bosr}[1]{\ensuremath{\mathit{D^{-}_{#1}}}\xspace}

\newcommand{\Polymax}{\ensuremath{\mathcal{P}_{\textsl{max}}}\xspace}
\newcommand{\Polymaxi}{\ensuremath{\mathcal{P}^i}\xspace}
\newcommand{\Polymaxiparam}[1]{\ensuremath{\mathcal{P}^{#1}}\xspace}

\newcommand{\Project}[2]{\ensuremath{\operatorname{\Pi}_{#1}}\big(#2\big)\xspace}
\newcommand{\Lift}[2]{\ensuremath{\operatorname{\Lambda}_{#1}}(#2)\xspace}

\newcommand{\hypro}{\textsc{HyPro}\xspace} 

\newcommand{\realyst}{\textsc{RealySt}\xspace}
\newcommand{\gsl}{\textsc{GNU Scientific Library}\xspace}
\newcommand{\prohver}{\textsc{ProHVer}\xspace}

\newcommand{\realystshort}{\textsc{R}}
\newcommand{\prohvershort}{\textsc{P}}

\newcommand{\scientific}[1]{\num[scientific-notation = true, exponent-product = \cdot]{#1}}
\newcommand{\samples}[1]{\scientific{#1}}
\newcommand{\errortable}[1]{\scriptsize\num[scientific-notation = true, exponent-product = \cdot, round-mode = places, round-precision = 3]{#1}}
\newcommand{\error}[1]{\num[scientific-notation = true, exponent-product = \cdot, round-mode = places, round-precision = 3]{#1}}
\newcommand{\probab}[1]{\num[round-mode = places, round-precision = 6]{#1}}

\newcommand{\intervals}[1]{\num{#1}}
\newcommand{\rt}[1]{\SI[round-mode = places, round-precision = 2]{#1}{\second}}

\newcommand{\tikzboxref}{%
    \begin{tikzpicture}[scale=0.2, baseline=0mm,  thick]
    \draw[refinement] (0,0) rectangle (1.4,0.8);
    \draw[cone,color2] (0,0) rectangle (1.4,0.8);
    \draw[border, color2] (0,0) rectangle (1.4,0.8);
    \end{tikzpicture}%
}
\newcommand{\tikzboxintermediateseg}{%
    \begin{tikzpicture}[scale=0.2, baseline=0mm,  thick]
    \draw[border, cinterseg, fill] (0,0) rectangle (1.4,0.8);
    \end{tikzpicture}%
}
\newcommand{\tikzboxgoal}{%
    \begin{tikzpicture}[scale=0.2, baseline=0mm,  thick]
    \draw[cone,boxcolor] (0,0) rectangle (1.4,0.8);
    \draw[border, boxcolor] (0,0) rectangle (1.4,0.8);
    \end{tikzpicture}%
}

\begin{document}

\title{Maximizing Reachability  Probabilities in Rectangular Automata with Random Clocks} %
\titlerunning{Reachability  Probabilities for Rectangular Automata with Random Clocks}

\author{Joanna Delicaris\inst{1}\orcidID{0000-0001-9455-4052} \and
Stefan Schupp\inst{2}\orcidID{0000-0002-2055-7581} \and
Erika \'{A}brah\'{a}m\inst{3}\orcidID{0000-0002-5647-6134} \and
Anne Remke\inst{1}\orcidID{0000-0002-5912-4767}}
\authorrunning{J. Delicaris et al.}
\institute{Westfälische Wilhelms-Universität, 48149 Münster, Germany 
\email{\{joanna.delicaris,anne.remke\}@uni-muenster.de}%
\and
TU Wien, Vienna, Austria\\
\email{stefan.schupp@tuwien.ac.at}
\and
RWTH Aachen, Aachen, Germany\\
\email{abraham@informatik.rwth-aachen.de}
}

\maketitle

\beginreview
\begin{abstract}
This paper proposes an algorithm to maximize reachability probabilities for %
rectangular automata with random clocks via a history-dependent prophetic scheduler. 
This model class incorporates time-induced nondeterminism on discrete behavior and nondeterminism in the dynamic behavior.
After computing reachable state sets via a forward flowpipe construction, we use backward refinement to compute maximum reachability probabilities.
The feasibility of the presented approach is illustrated on a  scalable model. 

\end{abstract}

\section{Introduction}
Hybrid automata \cite{alur1995AlgorithmicAnalysisHybrid} are a modeling formalism for systems whose evolution combines continuous dynamics interrupted by discrete steps. 
This work considers a subclass of rectangular automata \cite{henzinger2000TheoryHybridAutomata}, which we equip with stochasticity 
via random delays. %
The duration of a random delay in a hybrid automaton can be measured either (i) implicitly, via the semantics, or (ii) explicitly via a stopwatch and constraints on the syntax.
While the first is user-friendly and intuitive, the latter makes restrictions on the modeling formalism  explicit.

We follow the syntactical modeling variant and explicitly define the corresponding modeling restrictions. %
Similar to \cite{dargenio2005TheoryStochasticSystems,Pilch2021Optimizing}, we use stopwatches to model random delays on jumps.
We propose an algorithm to optimize reachability probabilities in  \emph{rectangular automata with random clocks} (\emph{RAR}). 

Nondeterminism, which arises naturally, e.g., in concurrent systems, is often resolved probabilistically in stochastic models \cite{Lygeros2010,AbateSHS2,bertrand2014StochasticTimedAutomata}, and is usually not explicitly resolved in non-stochastic hybrid systems. 
Recently, \emph{history-dependent} schedulers have been proposed to resolve \emph{discrete} nondeterminism in hybrid Petri nets \cite{pilch2020ClassicNonPropheticModel} and in singular automata with random clocks and urgent transitions either prophetically, or nonprophetically \cite{Pilch2021Optimizing}. 
The prophetic scheduler knows the  expiration times of random variables and  is considered more powerful than the nonprophetic scheduler, who does not know these. 
 Prophetic schedulers model the \emph{worst/best case scenario} and induce maximal bounds on probabilities.

When adding random clocks to rectangular automata, the challenge lies in the correct handling of continuous nondeterminism to compute correct probabilities.
We propose a measure-driven backward computation which partitions the infinite set of schedulers according to their ability to reach the goal set. 
Prophetic scheduling  hence computes a symbolic refinement of schedulers when performing a backward analysis through the precomputed reachability tree.  
Maximizing reachability probabilities requires taking the union over the reachable states leading to the goal, whose transition delays have been refined by backward analysis.  
To compute the optimal reachability probabilities, that union is projected  onto the integration domain of the corresponding probability densities, before performing a multidimensional integration over the joint probability density.

For a bounded number of jumps  reachability is decidable for non-initialized rectangular automata~\cite{alur1995AlgorithmicAnalysisHybrid,frehse2005PHAVerAlgorithmicVerification}. 
Hence,  flowpipe construction computes the exact reachable state-set for this model class, e.g. using a state representation of  polytopes~\cite{frehse2005PHAVerAlgorithmicVerification}.
Backward refinement is then used to resolve the inherent continuous nondeterminism such that  reachability probabilities are maximized. 
Consequently, the analysis approach presented here is  exact up to numerical integration for the considered model formalism.  %
To the best of our knowledge, the only approach able to compute reachability for this model class without resolving nondeterminism probabilistically computes a safe overapproximation via the tool \prohver~\cite{hahn2013CompositionalModellingAnalysis}. %
The feasibility of our approach is illustrated on a scalable model and validated by results computed in \prohver.

\paragraph{Related work.}

The application of model checking methods for probabilistic HA was enabled by CEGAR-style abstractions~\cite{zhang2012SafetyVerificationProbabilistic}. %
Extending decidable subclasses of HA by discrete probability distributions on jumps (\cite{sproston2000DecidableModelChecking},\cite{sproston2019Verification}) preserves decidability of reachability.
An extension of probabilistic timed automata is presented in~\cite{kwiatkowska2000VerifyingQuantitativeProperties}, where continuously distributed resets are used and randomized schedulers resolve \emph{discrete} nondeterminism for a discretized state-space.
Further approaches for (networks of) stochastic timed automata (\cite{ballarini2013TransientAnalysisNetworks}) maintain the probabilistic approach of resolving nondeterminism. 
Approaches for more general classes either rely on stochastic approximation (also for nondeterminism) ~\cite{prandini2006StochasticApproximationMethod} or on a combination of  discretization and randomness \cite{koutsoukos2008ComputationalMethodsVerification}. 
Several approaches that abstract to finite Markov decision processes have been proposed:
In~\cite{soudjani2015FAUST2FormalAbstractions} abstractions for uncountable-state discrete-time stochastic processes are proposed for SHA, where all nondeterminism is resolved probabilistically.
In~\cite{cauchi2019StocHyAutomatedVerification} an abstraction to interval Markov decision processes is proposed. Both approaches feature a stochastic kernel, and can hence not be compared to our work.

 The analysis proposed in~\cite{hahn2013CompositionalModellingAnalysis} resolves \emph{discrete} nondeterminism prophetically and  \emph{continuous} nondeterminism via a safe overapproximation (c.f. \cite{zhang2012SafetyVerificationProbabilistic}). 
 The same approach has been specified for stochastic timed automata in \cite{hahn2014ReachabilityRewardChecking}.

In~\cite{dargenio2018HierarchySchedulerClasses}, \emph{discrete} nondeterminism is resolved (non-)prophetically  for stochastic automata, %
where all continuous variables are random clocks. %
Similarly, scheduling of \emph{discrete} nondeterminism is introduced for hybrid Petri nets  in~\cite{pilch2020ClassicNonPropheticModel} and for singular automata with random clocks in \cite{Pilch2021Optimizing}, where a forward analysis without refinement is sufficient to compute maximal reachability probabilities.

\paragraph{Contribution.} 
We propose both
(i) the \emph{modeling formalism} of rectangular automata with random clocks, which combines discrete and continuous nondeterminism with stochastic delays; and (ii) an \emph{analytical approach} which computes maximum reachability probabilities induced by a prophetic scheduler. 
 We provide a feasibility study that shows that our computations are highly accurate.%

\paragraph{Organization of the paper.} Section~\ref{sec:modelclass} introduces the considered model class. The computation of the maximum probabilities is explained in Section~\ref{sec:computation}. The feasibility study is shown in Section~\ref{sec:casestudy} and the paper is concluded in Section~\ref{sec:conclusion}.

\section{Rectangular Automata with Random Clocks}\label{sec:modelclass}
Let $\mathbb{I}$ denote the set of all closed intervals $[a,b],[a,\infty),\allowbreak(-\infty,b],\allowbreak(-\infty,\infty)\allowbreak\subseteq\Reals$ with infinite or rational endpoints $a,b\in\Q$, with the standard set semantics. 

For $\CDF:\Realsposzero\rightarrow[0,1]\subseteq\Reals$ let $\support(\CDF)=\{v\in\Realsposzero\,|\,\CDF(v)>0\}$.
We call $\CDF$ a \emph{continuous distribution} if it is absolute continuous with $\int_{0}^{\infty}\CDF(v)dv=1$.
We call $\CDF$ a \emph{discrete distribution} if $\support(\CDF)$ is countable and $\sum_{v\in\support(\CDF)}f(v)=1$. We use $\CDFs$ and $\DDFs$ to denote the set of all continuous resp. discrete distributions, and let $\DFs=\CDFs\cup\DDFs$ contain all \emph{distributions}.

\emph{Hybrid automata} \cite{alur1995AlgorithmicAnalysisHybrid} are a modeling formalism for systems whose evolution combines continuous dynamics (flow) interrupted by discrete steps (jumps). 
In this work we restrict ourselves to the subclass of \emph{rectangular automata} \cite{henzinger2000TheoryHybridAutomata}.

\begin{definition}
    \label{def:RA}
    A \emph{rectangular automaton (RA)} is a tuple $\AR = \allowbreak(\Loc, \allowbreak\Var, \allowbreak\Inv,  \allowbreak\Init,\allowbreak\Act,\allowbreak\Edge)$ with 
    \begin{itemize}
        \item  a finite set $\Loc$ of \emph{locations}; 
    \item  a finite ordered set $\Var=\{x_1,{\ldots},x_{\dimC}\}$ of
    \emph{variables}; for $\val=(\val_1,{\ldots},\val_{\dimC}) \in \Reals^\dimC$ we define $\val_{x_i}=\val_i$, and for $I=I_1{\times}{\ldots}{\times} I_\dimC\in\mathbb{I}^\dimC$ we set $I_{x_i}=I_i$;
    \item  functions $\Inv$, $\Init$ and $\Act$, all of type $\Loc \rightarrow \mathbb{I}^{\dimC}$, assigning an \emph{invariant}, \emph{initial states} resp. \emph{flow} to each location; we call $\Act(\ell)_x$ the \emph{rate} of $x\in \Var$ in location $\ell\in \Loc$; 
    \item     a finite set  $\Edge \subseteq  \Loc \times \mathbb{I}^{\dimC} \times \mathbb{I}^{\dimC} \times 2^{\VarAut} \times \Loc$ of \emph{non-stochastic jumps} $e=(\ell,\pre ,\post ,\reset ,\ell')\in\Edge$ with \emph{source (target) location} $\ell$ ($\ell'$), such that $\pre_x=\post_x$ for all $x\in \Var\setminus\reset$ and $\post\subseteq\Inv(\ell')$; 
    \end{itemize}
$\AR$ is \emph{non-blocking} if for each location $\ell\in\Loc$ and each variable $x\in\Var$:
\begin{itemize}
\item if $\Inv(\ell)_x$ is lower-bounded by $a\in\Q$ (i.e. it has the form either $[a,b]$ or $[a,\infty)$) and $\Act(\ell)_x\cap\Realsneg\not=\emptyset$ then  there exists a non-stochastic jump $e= (\ell,\pre,\post,\reset,\ell')\in \Edge$ such that $\{\val\in\Inv(\ell) \mid \val_x=a\}\subseteq \pre$, 
\item if $\Inv(\ell)_x$ is upper-bounded by $b\in\Q$ (i.e. it has the form either $[a,b]$ or $(-\infty,b]$) and $\Act(\ell)_x\cap\Realspos\not=\emptyset$ then there exists a non-stochastic jump $e= (\ell,\pre,\post,\reset,\ell')\in \Edge$ such that $\{\val\in\Inv(\ell) \mid \val_x=b\}\subseteq \pre$.
\end{itemize}
\end{definition}

\begin{wrapfigure}[12]{l}{0.44\textwidth}%
\vspace*{-0.1cm}
\begin{tikzpicture}[
scale=1,
baseline,remember picture,
n/.style={draw, text width = 1.5cm, %
align = center, font= \footnotesize, rounded corners, very thick, execute at begin node=\setlength{\baselineskip}{8pt}%
},
en/.style={draw=none, minimum height=0cm, font = \scriptsize, align = center},%
c/.style={draw, fill, black, circle, inner sep=0, outer sep=0, minimum size=1mm},
l/.style={anchor=west, inner sep=0, font=\footnotesize},
]

\node[n, anchor=west](l0) at (-0.2,-1.7) {
	$\ell_0$\\
	\vspace{0.13cm} 	$\dot{x}=1$\\
	$\dot{y}\in [0,\nicefrac{1}{3}]$ \\	
	$\dot{r}=1$ \\ 
	\vspace{0.13cm} $x\leq 4$	};

\node[n, anchor=south](l1) at (2.5,-3.2) {$\ell_1$\\\vspace{0.13cm}
$\dot{x}=1$\\ 		
$\dot{y}=1$ \\	
$\dot{r}=0$ \\ \vspace{0.13cm}	
$y\leq 7$};

\node[n, anchor=south east](l2) at (5.2,-3.2) {$\ell_2$\\\vspace{0.13cm} 	
$\dot{x}=1$\\		
$\dot{y}=\nicefrac{1}{2}$ \\	
$\dot{r}=0$ };

\node[n, anchor=south
](l3) at (2.5,-1.2) {$\ell_3$\\\vspace{0.13cm} 
$\dot{x}=0$\\		
$\dot{y}=1$ \\	
$\dot{r}=0$  \\ \vspace{0.13cm}	
};

\node[n, anchor=south east](l4) at (5.2,-1.2) {$\ell_4$\\\vspace{0.13cm} 	
$\dot{x}=1$\\		
$\dot{y}=0$ \\	
$\dot{r}=0$ };

\node[en, text width=1.2cm,
anchor=south west, %
inner xsep=0pt
] (init) at 
(-0.2,-0.1)
{$x=1$\\$y\in[1,3]$\\$r=0$};

\draw[-latex, very thick] (init.south)  -- (l0.106);

\draw[-latex, very thick, bend left] (l0.70) to node[en, above, xshift=-0.1cm, yshift=0.1cm] {$r$}  (l3.190);

\draw[-latex, very thick, bend right] (l0.290) to node[en, left] {$x=4$} (l1.220);

\draw[-latex, very thick, bend left] (l1.40) to node[en, right, xshift=0.25cm] {$y\geq 5  $} (l2.110);

\draw[-latex, very thick, bend left] (l3.60) to node[en, above] {$x \geq 2,y\geq 10  $} (l4.120);

\end{tikzpicture}%
    \caption{RAR with $\VarRandom=\{r\}$.}\label{fig:runningex_automaton}%
\end{wrapfigure}
We equip non-blocking RA with stochastic delays modeled by \emph{random clocks}, which
behave as stopwatches, i.e. having derivatives either $0$ (\emph{paused}) or $1$ (\emph{active}). Each random clock $\clocki$ is associated with a continuous distribution to sample random variables $\randomvarij$, which determine at which value (\emph{expiration time}) of the random clock certain jumps shall be taken. After expiration, the random clock is reset to $0$ and a new random variable is used to determine the expiration time for the next event, yielding a sequence of random variables $\randomvar_{\clocki,0},\randomvar_{\clocki,1},\randomvar_{\clocki,2}, \dots$ for each random clock $r$. %
An illustrative example is depicted in Figure~\ref{fig:runningex_automaton}.

\begin{definition}
    \label{def:RHA}
    A \emph{rectangular automaton with random clocks (RAR)} is a tuple $\A = \allowbreak(\Loc, \allowbreak\Var, \allowbreak\VarRandom,\allowbreak\Distr, \allowbreak\Inv,  \allowbreak\Init,\allowbreak\Act, \allowbreak\ActRandom,\allowbreak\Edge,\allowbreak\EdgeRandom)$ with $(\Loc, \allowbreak\Var, \allowbreak\Inv,  \allowbreak\Init,\allowbreak\Act,\allowbreak\Edge)$ a non-blocking RA and:
    \begin{itemize}
     \item a finite ordered set $\VarRandom=\{\clocki_1,\ldots,\clocki_{\dimR}\}$ of \emph{random clocks}; we use analogously $\valRandom_{\clocki_i}=\valRandom_i$ for $\valRandom\in\Reals^{\dimR}$ and $I_{\clocki_i}=I_i$ for $I=I_1\times\ldots\times I_\dimR\in\mathbb{I}^\dimR$;
    \item  a function $\Distr:\VarRandom\rightarrow\CDFs$;
    \item a function $\ActRandom: \Loc \rightarrow \{0,1\}^{|\VarRandom|}$;
    \item  a finite set $\EdgeRandom \subseteq \Loc \times \VarRandom \times \Loc$ of \emph{stochastic jumps} $e=(\ell,r,\ell')$ with \emph{source (target) location} $\ell$ ($\ell'$) and random clock $r$, where 
    (i) each two stochastic jumps $e,e'\in\EdgeRandom$, $e\not =e'$, with the same source location $\ell$ have different random clocks,
    (ii)  for all locations $\ell \in \Loc$ and all random clocks $\clocki\in \VarRandom$, if $\ActRandom(\ell)_{\clocki} = 1$ then $(\ell,\clocki,\ell')\in \EdgeRandom$ for some $\ell'\in\Loc$, and
    (iii) for each stochastic jump $(\ell,r,\ell')\in \EdgeRandom$ it holds that $\Inv(\ell)\subseteq \Inv(\ell')$.
    \end{itemize}
\end{definition}

\noindent Let $\A = (\Loc, \allowbreak\Var, \allowbreak\VarRandom,\allowbreak\CDF, \allowbreak\Inv,\allowbreak  \Init,\allowbreak\Act,\allowbreak\Edge,\allowbreak \EdgeRandom)$ with $\Var=\{x_1,\ldots,x_{\dimC}\}$ and  $\VarRandom=\{\clocki_1,\ldots,\clocki_{\dimR}\}$ be a RAR.
A \emph{state} $\hastate=(\ell,\val, \valRandom, \sample)\in\States=\Loc\times\Reals^\dimC\times\Reals^\dimR\times\Reals^\dimR$ of $\A$ specifies the current location $\ell$, the values $\val$ of the continuous variables, and the values $\valRandom$ and expiration times $\sample$ of the random clocks.
 The  operational semantics \cite{henzinger1998WhatDecidableHybrid} specifies the evolution of a state of $\A$,  by letting time elapse or by taking a discrete jump (non-stochastic or stochastic):
\noindent
\begin{eqnarray*}
    & \inference[{Flow}]{
    t\in\Realsposzero \myquad
    \textit{rate} \in  \Act(\ell) \myquad
    \val'=\val+t\cdot \textit{rate} \myquad
    \val' \in \Inv(\ell) \\
\valRandom'=\valRandom+t \cdot \ActRandom(\ell) \myquad \forall \clocki \in \VarRandom.\, \valRandom'_\clocki \leq \sample_r
}
{
    (\ell,\val, \valRandom, \sample) \xrightarrow{t,\textit{rate}} (\ell,\val',\valRandom', \sample)
}
\end{eqnarray*}

\noindent
\begin{eqnarray*}
& \inference[Jump$_C$]{
    e=(\ell,\pre , \post , \reset , \ell') \in \Edge \myquad
     \val \in \pre \myquad \val' \in \post \\
     \forall x \in \Var\setminus\reset .\ \val'_{x}=\val_{x}\myquad \val' \in \Inv(\ell')
    }{
     (\ell,\val, \valRandom, \sample) \xrightarrow{e} (\ell',\val',\valRandom, \sample)
    }
\end{eqnarray*}

\noindent
\begin{eqnarray*}
& \inference[Jump$_R$]{
    e=(\ell,\clocki, \ell') \in \EdgeRandom \myquad \valRandom_\clocki = \sample_\clocki \myquad \valRandom'_\clocki = 0 \myquad
    \sample'_\clocki \in \support(\Distr({\clocki})) \\ 
    \forall \clocki' \in \VarRandom \setminus \{\clocki\}. 
    \valRandom'_{\clocki'}=\valRandom_{\clocki'} \land \sample'_{\clocki'} = \sample_{\clocki'}
    }{
     (\ell,\val, \valRandom, \sample) \xrightarrow{e} (\ell',\val,\valRandom', \sample')
    }
\end{eqnarray*}

For $\hastate\in\States$, let $\EnabledJumps(\hastate)=\{e\in\Edge\cup\EdgeRandom\,|\,\exists \hastate'\in\States.\,\hastate\xrightarrow{e}\hastate'\}$ be the set of jumps \emph{enabled} in $\hastate\in\States$.
We will also use
$\Time(\hastate):=\{(t,\mathit{rate})\in \Realsposzero\times \Reals^\dimC\,|\,\exists \hastate'\in\States.\,\hastate\xrightarrow{t,\textit{rate}}\hastate' \}$.

We set $\rightarrow = (\bigcup_{t\in\Realsposzero,\textit{rate}\in\Reals^\dimC}\xrightarrow{t,\textit{rate}})\cup(\bigcup_{e\in\Edge\cup\EdgeRandom}\xrightarrow{e})$. 
A \emph{path} of $\A$ is a finite sequence $\pi=\hastate_0\xrightarrow{a_0}\hastate_1\xrightarrow{a_1}\ldots$ of states with $\hastate_0=(\ell_0,\val_0, \valRandom_0, \sample_0)$, $\val_0\in\Inv(\ell_0)$, and $\hastate_i\xrightarrow{a_i}\hastate_{i+1}$ for all $i\in\Naturals$;
we call $\pi$ \emph{initial} if $\val_0\in\Init(\ell_0)$, $\valRandom_0 = 0 \in \Reals^\dimR$ and $\sample_{0_\clocki} \in \support(\Distr(\clocki))$ for all $\clocki \in \VarRandom$. %
A state is \emph{reachable} if there is an initial path leading to it.
Let $\Paths$ denote the set of all paths. %

For every reachable state, there is a path with alternating delays and jumps, as delays may have  duration zero and  consecutive delays can be combined.
This holds even for  consecutive delays with different rates, as flow sets are convex\footnote{For a proof of Lemma~\ref{lemma:rates}, we refer to~\ref{appendix:proofs}.}:

\begin{lemma}\label{lemma:rates}
Let $\hastate_0\xrightarrow{t_1,\textit{rate}_1}\hastate_1\xrightarrow{t_2,\textit{rate}_2}\hastate_2$ for $\hastate_0,\hastate_1,\hastate_2\in\States$ with location $\ell$, $t_1,t_2\in\Realsposzero$  and $\textit{rate}_1,\textit{rate}_2 \in \Act(\ell)$. Then there is  $\textit{rate}\in \Act(\ell)$ s.t. $\hastate_0\xrightarrow{t_1 + t_2,\textit{rate}}\hastate_2$.
\end{lemma} %

The \emph{duration} of  a path is defined as the sum of the durations of its steps, where jumps are considered instantaneous: $\dur(\hastate)=0$, $\dur(\pi\xrightarrow{e}\hastate)=\dur(\pi)$ and $\dur(\pi\xrightarrow{t,\textit{rate}}\hastate')=\dur(\pi)+t$. 
Let the \emph{jump-depth} of a path be the number of jumps in it. 
We call a RAR \emph{Zeno-free} iff for all $t\in\Realsposzero$ there exists a $k\in\Naturals$ such that all initial paths of jump-depth at least $k$ have duration at least $t$. 
We deviate from the standard definition, which requires that only finitely many jumps are possible in finite time. Our definition assures a  \emph{concrete bound on the number of jumps} per path for each time bound.
In the following, we assume all considered models to be Zeno-free.

RAR allow for (i) \emph{initial nondeterminism} in the choice of the initial state, (ii) \emph{time nondeterminism} when time can elapse but also jumps are enabled during the whole flow, and (iii) \textit{rate nondeterminism} when continuous variables can evolve with different rates. 
In addition to these continuous types of nondeterminism, we also consider (iv) \emph{discrete nondeterminism} when different jumps are enabled simultaneously.
We use  prophetic schedulers to resolve nondeterminism, which have full information not only on the history but also on the future expiration times of all random clocks, as introduced in \cite{Pilch2021Optimizing}.
While prophetic scheduling may seem unrealistic, they are well-suited to perform a \emph{worst-case} analysis, especially when uncontrollable uncertainties are modeled nondeterministically.

 \begin{definition}
 \label{def:Scheduler}
\sloppy A \emph{(prophetic history-dependent) scheduler} is a function  $\allowbreak\mathfrak{s} :\allowbreak \Paths \allowbreak\rightarrow \DFs$ which assigns 
to every
path $\Path =\hastate_0\xrightarrow{a_1}\ldots\xrightarrow{a_{n}}\hastate_n\in \Paths$
a distribution $\CDF=\mathfrak{s}(\Path)$,
such that 
if $n\geq 1$ and  $a_n \in \Edge\cup\EdgeRandom$ is an edge then $\CDF$ 
 is continuous with $\support(\CDF) \subseteq \Time(\hastate_n)$ and
otherwise $\CDF$ 
is discrete with $\support(\CDF)\subseteq  \EnabledJumps(\hastate_n)$.
The set of  schedulers is denoted $\SchedulersProphetic$.
 \end{definition} %
 
We  maximize the probability  of \emph{time-bounded} reachability, i.e., of reaching a  set of goal states $\Goal \subseteq \States$ along initial paths  $\pi$ with $\dur(\hahistory)\leq \tmax$. %

 Resolving nondeterminism in a  RAR $\A$ via the set of schedulers $\SchedulersProphetic$ induces an interval of \emph{reachability probabilities} $[p^{\SchedulersProphetic}_\textsl{min}(\Goal, \tmax)$,\allowbreak $p^{\SchedulersProphetic}_\textsl{max}(\Goal, \tmax)]$, where the bounds are referred to as minimum and maximum. 
We define $\SchedulersProphMax \subseteq\SchedulersProphetic$ as the set of schedulers that reach $\Goal$ along initial paths $\hahistory$ with $\dur(\hahistory)\leq \tmax$ and  induce $p^{\SchedulersProphetic}_\textsl{max}$.
Let $\valSet^\scheduler \subseteq \Reals^\dimR$ denote the sample values for all random variables that allow scheduler $\scheduler$ to reach $\Goal$.  
This yields the following definition:

\begin{definition}
    \label{lemma:probability} 
 The prophetic maximum reachability probability is:
\begin{equation}\label{eq:scheduler}
p^{\SchedulersProphetic}_\textsl{max}(\Goal, \tmax) 
= \int_{\bigcup_{\scheduler \in \SchedulersProphMax}\valSet^\scheduler} G(\assignment) \ d\assignment
,
\end{equation}
where  $G(\assignment)= \prod_{\contrandomvar} \Distr(\clocki) $ is the joint probability density function for all random delays $r_n$ and random clocks $r$. %
\end{definition}
Note that due to the independence of the random variables, $G(\assignment)$ equals  the product over the probability density functions $ \Distr(\clocki)$.

\section{Computation of Maximum Reachability Probabilities }\label{sec:computation}

The inherent continuous nondeterminism in RAR leads to an   uncountable number of choices and hence  schedulers. 
We propose a measure-driven state space construction, which partitions the infinite set of continuous schedulers w.r.t. their ability to reach the set of goal states. 
We remark that  our model class allows resets on continuous variables, but our method does not yet  support this. %

Section~\ref{subsec:forward} explains the forward flowpipe construction and Section~\ref{subsec:refinement} introduces the backward refinement.   
Sample domains are extracted  in Section~\ref{subsec:collection} and  maximum prophetic reachability probabilities are computed in Section~\ref{subsec:schedulers}.

\subsection{Forward flowpipe construction}\label{subsec:forward}

To compute reachability for rectangular automata, flowpipe construction has been proposed in \cite{alur1995AlgorithmicAnalysisHybrid}, 
which results in a geometric representation of all states reachable up to a predefined time bound.
To apply this method to RAR, we disregard the stochasticity by removing all constraints on the sample values $\sample$ from the operational semantics.
The resulting automaton is a regular rectangular automaton, where the $n$-th random delay induced by random clock $r$ in the original automaton is treated as a continuous variable $\contrandomvar$. 
Replacing every random clock $\clocki$ with one continuous variable for each possible delay corresponds to an enrollment of the automaton (c.f. Figure~\ref{fig:enrollment}). %
The set $\VarRandom$ of the enrolled rectangular automaton then contains continuous variables $\clocki_0, \clocki_1, \dots$ for each random clock $r$ in the original automaton and $\dimR$ is the number of all random delays.

In the following, we omit the sampled values $\sample$ from states $\hastate = (\ell, \val, \valRandom, \sample)$. 
To simplify, we also combine the valuations of continuous variables $\val$ and random clocks $\valRandom$, such that we call $(\ell, \valSet)$ \emph{state set}, where $\valSet$ is a set of valuations and $\val_x$ refers to the valuation of a variable $x \in \Var \cup \VarRandom$.
Even though formally we work on state sets $(\ell, \valSet)$, for readability our notation restricts to valuation sets $\valSet$. 
We refer to valuation sets as \emph{segments} and to the respective location of the state set as \emph{corresponding} location.

\begin{figure}[t]
    \centering
        \begin{minipage}[b]{.35\linewidth}
        \centering
           \begin{tikzpicture}[
scale=1,
node distance = 0.4cm and 0.25cm,
baseline,remember picture,
n/.style={draw, text width = 1.2cm, %
align = center, font= \footnotesize, rounded corners, very thick, execute at begin node=\setlength{\baselineskip}{8pt}%
},
en/.style={draw=none, minimum height=0cm, font = \scriptsize, align = center},%
c/.style={draw, fill, black, circle, inner sep=0, outer sep=0, minimum size=1mm},
l/.style={anchor=west, inner sep=0, font=\footnotesize},
edge/.style={-latex, very thick},
]

\node[n, anchor=north] (l) at (0,0) {
$\dot{r}=1$
};
\node[n, below=0.8cm of l] (lt)  {
$\dot{x}=2$
};

\node[en,above=0.7cm of l] (initl) {};
\draw[edge] (initl) -- node[en, left] {$r=0$\\$x=0$} (l);
\draw[edge] (l.300) to node[en,right] {$r$}  (lt.60);
\draw[edge] (lt.120) to node[en,left] {$x=2$\\$x:=0$}  (l.240);

\node[n, anchor=north] (l0) at (1.8,0) {
$\dot{r_0}=1$
};
\node[n,  below= of l0] (l0t) {
$\dot{x}=2$
};
\node[n, below=of l0t] (l1) {
$\dot{r_1}=1$
};
\node[n, below=of l1] (l1t) {
$\dot{x}=2$
};

\node[en,above= 0.7cm of l0] (initl0) {};

\draw[edge] (initl0) -- node[en, right] {$r_0=0$\\$r_1=0$} (l0);
\draw[edge] (initl0) -- node[en, left] {$x=0$} (l0);

\draw[edge] (l0) to node[en, right] {$r_0$} (l0t);
\draw[edge] (l1) to node[en, right] {$r_1$} (l1t);
\draw[edge] (l0t) to node[en,left] {$x=2$}  (l1);
\draw[edge] (l0t) to node[en,right] {$x:=0$}  (l1);

\end{tikzpicture}
           \caption{Enrollment of RAR with $\VarRandom=\{r\},\tmax=3$; rates are zero if not stated.}\label{fig:enrollment}
        \end{minipage}\hfill%
        \begin{minipage}[b]{.6\linewidth}
       \begin{subfigure}[b]{.48\linewidth}
           \definecolor{solBase01}{RGB}{88 110 117}
\definecolor{mainColor}{named}{solBase01}

\definecolor{solMagenta}{RGB}{211  54 130}
\definecolor{color2}{named}{solMagenta}
\definecolor{solGreen}{RGB}{133 153   0}
\definecolor{color3}{named}{solGreen}
\definecolor{solYellow}{RGB}{181 137   0}
\definecolor{color4}{named}{solYellow}

\definecolor{solBlue}{RGB}{38 139 210}
\definecolor{color6}{named}{solBlue}

\definecolor{dartmouthgreen}{rgb}{0.05, 0.5, 0.06}
\definecolor{color5}{named}{dartmouthgreen}

\definecolor{davysgrey}{rgb}{0.33, 0.33, 0.33}
\definecolor{boxcolor}{named}{color4}

\definecolor{fancyblue}{rgb}{0.01, 0.28, 1.0}

\definecolor{cinitloc}{named}{mainColor}
\definecolor{cpath1light}{named}{color6}
\definecolor{cpath1dark}{named}{fancyblue}
\definecolor{cpath2light}{named}{color3}
\definecolor{cpath2dark}{named}{color5}

\tikzstyle{initloc}=[cinitloc]
\tikzstyle{path1start}=[cpath1light,densely dotted]
\tikzstyle{path1end}=[cpath1dark]
\tikzstyle{path2start}=[cpath2light,dotted]
\tikzstyle{path2end}=[cpath2dark]

\tikzstyle{filling}=[very thick, fill, fill opacity = 0.2]
\tikzstyle{cone}=[draw = none, fill, fill opacity = 0.2]
\tikzstyle{border}=[very thick, line join = round]
\tikzstyle{guard}=[fill opacity=0.1]
\tikzstyle{guardticks}=[color3]
\tikzstyle{init}=[border, mainColor, line join = round]

\begin{tikzpicture}[
baseline,remember picture,
font=\footnotesize, 
tick/.style={fill=white,font= \scriptsize}
]

\useasboundingbox (-0.75,-0.375) rectangle (3,3.4);

\node (ursprung) at (0,0) {};
\draw[help lines, lightgray, xstep=0.25, ystep=0.25] ($(ursprung)-(0.125,0.125)$) grid (2.875, 3.125);
\node[fill=white] (y) at (0,3) {$y$};
\node[fill=white] (x) at (2.75,0) {$x$};

\foreach \y in {2,4,...,10}  {
    \node[tick] (\y) at (-0.25,0.25*\y) {$\y$}; 
    \draw (-0.05,0.25*\y)--(+0.05,0.25*\y);
}
\foreach \x in {2,4,...,8}  {
    \node[tick] (\x) at (0.25*\x,-0.25) {$\x$}; 
    \draw (0.25*\x,-0.05)--(0.25*\x,+0.05);
}

\draw[-latex,thick] ($(ursprung)-(0,0.125)$) to (y);
\draw[-latex,thick] ($(ursprung)-(0.125,0)$) to (x);

\coordinate (init-l) at (0.25,0.25);
\coordinate (init-u) at (0.25,0.75);
\coordinate (s1-1-l) at (1,0.25);
\coordinate (s1-1-u) at (1,1);
\coordinate (s1-2-l) at (2.5,1.75);
\coordinate (s1-2-u) at (1.75,1.75);
\coordinate (s2-1-l) at (2,1.25);
\coordinate (s2-1-u) at (1.25,1.25);
\coordinate (s2-2-l) at (2.75,1.625);
\coordinate (s2-2-u) at (2.75,2.25);
\coordinate (s3-1-l) at ($(init-l)$);
\coordinate (s3-1-u) at (0.25,3);
\coordinate (s3-2-l) at ($(s1-1-l)$);
\coordinate (s3-2-u) at (1,3);
\coordinate (s4-1-l) at (0.5,2.5);
\coordinate (s4-1-u) at (0.5,3);
\coordinate (s4-2-l) at (2.75,2.5);
\coordinate (s4-2-u) at (2.75,3);

\draw[init] ($(init-l)-(0,0)$) -- ($(init-u)-(0,0)$) -- ($(init-u)-(0.03,0)$) -- ($(init-l)-(0.03,0)$) -- cycle;

\draw[cone, mainColor] (init-l) -- (s1-1-l) -- (s1-1-u) -- (init-u) -- cycle;
\draw[cone, mainColor] (s1-1-l) -- (s1-1-u) -- (s1-2-u) -- (s1-2-l) -- cycle;
\draw[cone, mainColor] (s2-1-l) -- (s2-1-u) -- (s1-2-u) -- (s2-2-u) -- (s2-2-l) -- cycle;

\draw[border, initloc] (init-l) -- ($(s1-1-l)-(0.0,0)$) -- ($(s1-1-u)-(0.0,0)$) -- (init-u) -- cycle;
\draw[border, path1end] (s2-2-l) -- (s2-1-l) -- (s2-1-u) -- (s1-2-u) -- (s2-2-u) ; %
\draw[border, path1start] (s1-1-l) -- (s1-1-u) -- (s1-2-u) -- (s1-2-l) -- cycle;

\draw[cone, mainColor](s3-2-u) -- (s3-2-l)-- (s3-1-l) -- (s3-1-u);
\draw[cone, mainColor](s4-2-u) -- (s4-2-l)-- (s4-1-l) -- (s4-1-u);

\draw[border,path2start ](s3-2-u) -- (s3-2-l)-- (s3-1-l) -- (s3-1-u); %
\draw[border,path2end ] (s4-2-l)-- (s4-1-l) -- (s4-1-u); %

\draw[border, boxcolor] (2,2) rectangle (2.5,2.75);
\draw[cone, boxcolor] (2,2) rectangle (2.5,2.75);

\end{tikzpicture}
           \caption{Flowpipe and goal.}\label{fig:runningex_flowpipe}
        \end{subfigure}\hfill%
        \begin{subfigure}[b]{.48\linewidth}
           \begin{tikzpicture}[
scale=1,
node distance = 0.4cm and 0.2cm,
baseline,
remember picture,
n/.style={draw, text width = 0.8cm, text = black, %
align = center, font= \footnotesize, rounded corners, very thick, execute at begin node=\setlength{\baselineskip}{8pt}%
},
en/.style={draw=none, minimum height=0cm, font = \scriptsize, align = center},%
ref/.style={draw, circle, minimum width=1.5mm, inner sep=0, fill,refinementcolor},%
c/.style={draw, fill, black, circle, inner sep=0, outer sep=0, minimum size=1mm},
l/.style={anchor=west, inner sep=0, font=\footnotesize},
]

\useasboundingbox (-2,0.25) rectangle (2,-3);

\node[n, draw=cinitloc](l0) at (0,0) {	$\ell_0$	};
\node[n, draw=cpath1light,below left = of l0.south]  (l1)  {$\ell_1$};
\node[n, draw=cpath1dark, below = of l1] (l2)  {$\ell_2$};
\node[n, draw=cpath2light, below right = of l0.south] (l3) {$\ell_3$};
\node[n, draw=cpath2dark, below = of l3] (l4) {$\ell_4$ };

 \draw[-latex, very thick] (l0) to node[en, left] {}  (l1);
 \draw[-latex, very thick] (l0) to node[en, above] {} (l3);
 \draw[-latex, very thick] (l1) to node[en, left] {} (l2);
 \draw[-latex, very thick] (l3) to node[en, left] {} (l4);

\end{tikzpicture}
           \caption{Reach tree.}\label{fig:runningex_reachtree}
        \end{subfigure}
        \caption{Flowpipe (in 2D), goal set and an illustration of the reach tree for the RAR shown in Figure~\ref{fig:runningex_automaton}.}
    \label{fig:runningex_forward}
        \end{minipage}
\end{figure}

We  execute a forward flowpipe construction, i.e., starting with the initial  states we alternate between computing the forward time closure and the jump successors until a predefined $\tmax$ is reached.
The forward time closure $\ftc{\ell}(\valSet)$ of $\valSet$ in $\ell$ (c.f.  \cite{alur1995AlgorithmicAnalysisHybrid}), represents the set of states reachable from the set of entry states, \ie computes states reachable via a time delay in  location $\ell$. 
Jumps are represented by the one-step relation $\fosr{e}(\valSet)$, which defines the set of valuations reachable by choosing transition $e\in \Edge\cup\EdgeRandom$ from  valuations $\valSet$~\cite{alur1995AlgorithmicAnalysisHybrid}.

Our computation relies on a state representation via convex polytopes, usually in $\hrep$, defined as an intersection of multiple halfspaces. 
Some computations require the $\vrep$, defined as  convex hull of a set of vertices (convex hull combined with convex cone for unbounded polytopes) \cite{Ziegler1995}.

The forward flowpipe construction
then computes all reachable segments and stores them together with their corresponding location in the \emph{reach tree} $\reachtree$ according to their occurrence.
We define a \emph{reach tree}: %
\begin{definition}
    For a rectangular automaton $R$ and a time bound $\tmax$, 
    a \emph{reach tree} is a tuple $\reachtree = (N,E)$ with symbolic state sets $(\ell,\valSet)\in N$ as nodes and edges $e\in E = N \times  (\Edge \cup \EdgeRandom) \times N$ annotated with jumps, 
    where for each state $\hastate=(\ell,\val)\in \States$ it holds that, if and only if $\hastate$ is reachable in $R$ before time $\tmax$, there exists a node $n=(\ell,\valSet)\in N$, such that $\sigma \in \valSet$.
\end{definition}

When performing qualitative analysis in the purely hybrid case, flowpipe construction ends as soon as the intersection of a computed segment with the set of goal states is non-empty. 
As we perform quantitative reachability analysis, we have to complete the flowpipe construction until the time bound $\tmax$ and collect all trajectories leading to goal states in order to fully resolve the nondeterminism present in the system.
After computing the flowpipe  up to  $\tmax$, the resulting segments are intersected with the goal set to determine  reachability.
We define goal states as $\Goal = (\Goallocations, \Goalvalset)$, where $\Goallocations$ is a set of locations and $\Goalvalset$ is a set of valuations defined as a convex polytope, with constraints only for continuous variables $x\in \Var$.
Hence, $\Goalvalset$ is unbounded in the dimensions of random clocks $r \in \VarRandom$. %

    We define the set $\Goalreached$ of \emph{reachable goal states}, such that it contains all subsets of the goal state set that can be reached via a trace in the reach tree:
    \begin{align*}
        (\ell,\valSet) \in \Goalreached \Leftrightarrow \ell \in \Goallocations \land \exists (\ell',\valSet') \in \reachtree. \valSet' \cap \Goalvalset = \valSet (\not = \emptyset) \land \ell'=\ell.
    \end{align*}
    We use $\Goalvali$  and $\Goali=(\ell,\Goalvali)$ to refer to these subsets of goal states and index $i$ to refer to the trace of the reach tree leading to $\Goali$ (c.f. Figure~\ref{fig:targettree}). 
    We define $\pathindices$ as the set collecting all trace indices $i$.
All segments $\Goalvali$, then  serve as input for the backward refinement (c.f.  Section~\ref{subsec:refinement}). 

\begin{runningexample}
The forward time closure of the initial set in  location $\ell_{0}$ (c.f. Figure~\ref{fig:runningex_automaton}) corresponds to the segment indicated by a gray solid line  in Figure~\ref{fig:runningex_flowpipe}.
Moving from location $\ell_0$ to $\ell_3$ corresponds to the expiration of the random clock $r$, which models the only random delay present in the automaton.
In location $\ell_2$, only $y$ evolves (green dotted border). %
For states with $x\geq2$ and $y\geq 10$, taking the transition to location $\ell_4$ is possible (solid dark green border).
Here, $y$ is stopped and only $x$ evolves. %
Alternatively,  moving from $\ell_0$ to $\ell_1$ (blue dotted border) is possible for $x=4$.
All states with $y\geq 5$ can reach location $\ell_2$,  as long as $y\leq 7$ holds.
This leads to overlapping segments for $\ell_1$ and $\ell_2$ (solid dark blue border).
The goal states (yellow border) are defined by $\Goal =\{(\ell,\val) \in \States \mid \ell \in Loc \;\land\; \val_x \in [8,10] \land \val_y \in [8,11] \}$. %
  Figure~\ref{fig:runningex_reachtree} illustrates the reach tree $\reachtree$.

  The goal can be reached both, via locations $\ell_1,\ell_2$ and via locations $\ell_3,\ell_4$.
  Hence, intersecting the forward flowpipe segments with the goal set results in two traces $i=0,1$, leading to state sets $\States^0=(\ell_2,\valSet^0)$ and $\States^1=(\ell_4,\valSet^1)$, where:
 \begin{align*}
 \valSet^0 &= \{ \val \in \mathbb{R}^3 \mid \val_x \leq 10 \land \val_y \geq 8 \land \val_y \leq 3.5+\nicefrac{1}{2}\cdot \val_x \land \val_r=3\}\\      \valSet^1 &= \{ \val \in \mathbb{R}^3 \mid \val_x \in [8,10] \land \val_y \in [10,11] \land  \val_r\in [0,3]\}.
 \end{align*}

\end{runningexample}

\subsection{Backward refinement}\label{subsec:refinement}

Starting from each $\Goalvali\subseteq\Goalvalset$, 
we  perform a  backward computation along the reach tree to refine  state sets according to prophetic maximum schedulers until the root of the tree is reached. 
We refine backward by computing \emph{refined segments} and \emph{intermediate goal segments}  for all forward flowpipe segments on trace $i$. 
The result of the backward refinement then is given by all refined segments and $\Goalvali$ in traces $i$ along the reach tree, containing exactly the fragment of the reach tree that allows reaching the goal.

Backward propagation relies on the definitions of backward time closure $\btc{\ell}(\valSet)$ and backward one-step relation $\bosr{e}(\valSet)$ (c.f.~\cite{alur1995AlgorithmicAnalysisHybrid}). 
Similar to the forward time closure, $\btc{\ell}(\valSet)$ computes all states valid in location $\ell$ by regressing the time delay in that location, that are able to reach valuations in $\valSet$.
$\bosr{e}(\valSet)$ reverts the effects of transition $e$ leading to $\valSet$, defined through guards and resets.

Starting from each segment $\Goalvali$, the backward time closure $\btc{\ell}(\valSet)$ is computed according to the activity of the corresponding location $\ell$.
This yields an unbounded cone, containing all states which can reach $\Goalvalset$ from an arbitrary state in location $\ell$. 
We define the corresponding precomputed forward segment $\Goalvalik{k}$
as the flowpipe segment on trace $i$ 
in the reach tree, which is encountered in step $k$ going backward from $\Goalvali$ for $\Goalvali \subset \Goalvalik{0}$ (c.f. Figure~\ref{fig:runningex_refinement_names}).
It is then intersected with the unbounded backward time closure to restrict the state set to states that are actually reachable in the hybrid automaton via a maximizing scheduler (c.f. Figure~\ref{fig:backward_ref_steps0}). 
This results in a so-called \emph{refined segment} $\Refvalik{k}$, containing all states which can reach the goal set from location $\ell$ (in step $k$) on trace $i$:
\begin{figure}[t]
    \centering
       \begin{subfigure}[b]{.28\linewidth}
       \centering
        \definecolor{solBase01}{RGB}{88 110 117}
\definecolor{solMagenta}{RGB}{211  54 130}
\definecolor{color2}{named}{solMagenta}
\definecolor{solGreen}{RGB}{133 153   0}
\definecolor{color3}{named}{solGreen}
\definecolor{solYellow}{RGB}{181 137   0}
\definecolor{color4}{named}{solYellow}
\definecolor{dartmouthgreen}{rgb}{0.05, 0.5, 0.06}
\definecolor{color5}{named}{dartmouthgreen}
\definecolor{solBlue}{RGB}{38 139 210}
\definecolor{color6}{named}{solBlue}
\definecolor{davysgrey}{rgb}{0.33, 0.33, 0.33}
\definecolor{fancyblue}{rgb}{0.01, 0.28, 1.0}

\definecolor{mainColor}{named}{solBase01}
\definecolor{boxcolor}{named}{color4}
\definecolor{cinitloc}{named}{mainColor}
\definecolor{cpath1light}{named}{color6}
\definecolor{cpath1dark}{named}{fancyblue}
\definecolor{cpath2light}{named}{color3}
\definecolor{cpath2dark}{named}{color5}

\tikzstyle{initloc}=[cinitloc]
\tikzstyle{path1start}=[cpath1light,densely dotted]
\tikzstyle{path1end}=[cpath1dark]
\tikzstyle{path2start}=[cpath2light,dotted]
\tikzstyle{path2end}=[cpath2dark]

\tikzstyle{filling}=[very thick, fill, fill opacity = 0.2]
\tikzstyle{cone}=[draw = none, fill, fill opacity = 0.2]
\tikzstyle{border}=[very thick, line join = round]
\tikzstyle{guard}=[fill opacity=0.1]
\tikzstyle{guardticks}=[color3]
\tikzstyle{init}=[border, mainColor, line join = round]
\tikzstyle{refinement}=[pattern=north west lines, pattern color = color2]
\tikzstyle{refinement2}=[pattern=north east lines, pattern color = color2]

\begin{tikzpicture}[
baseline, remember picture,
font=\footnotesize, scale=1, 
tick/.style={fill=white,font= \scriptsize}]

\useasboundingbox (-0.25,-0.25) rectangle (3,3);

\node (ursprung) at (0,0) {};
\draw[help lines, lightgray, xstep=0.25, ystep=0.25] ($(ursprung)-(0.125,0.125)$) grid (2.875, 3.125);
\node[fill=white] (y) at (0,3) {$y$};
\node[fill=white] (x) at (2.75,0) {$x$};

\foreach \y in {2,4,...,10}  {
    \node[tick] (\y) at (-0.25,0.25*\y) {$\y$}; 
    \draw (-0.05,0.25*\y)--(+0.05,0.25*\y);
}
\foreach \x in {2,4,...,8}  {
    \node[tick] (\x) at (0.25*\x,-0.25) {$\x$}; 
    \draw (0.25*\x,-0.05)--(0.25*\x,+0.05);
}

\draw[-latex,thick] ($(ursprung)-(0,0.125)$) to (y);
\draw[-latex,thick] ($(ursprung)-(0.125,0)$) to (x);

\coordinate (goal-l) at (2,2);
\coordinate (goal-u) at (2.5,2.75);
\coordinate (init-l) at (0.25,0.25);
\coordinate (init-u) at (0.25,0.75);
\coordinate (s1-1-l) at (1,0.25);
\coordinate (s1-1-u) at (1,1);
\coordinate (s1-2-l) at (2.5,1.75);
\coordinate (s1-2-u) at (1.75,1.75);
\coordinate (s2-1-l) at (2,1.25);
\coordinate (s2-1-u) at (1.25,1.25);
\coordinate (s2-2-l) at (2.75,1.625);
\coordinate (s2-2-u) at (2.75,2.25);
\coordinate (s3-1-l) at ($(init-l)$);
\coordinate (s3-1-u) at (0.25,3);
\coordinate (s3-2-l) at ($(s1-1-l)$);
\coordinate (s3-2-u) at (1,3);
\coordinate (s4-1-l) at (0.5,2.5);
\coordinate (s4-1-u) at (0.5,3);
\coordinate (s4-2-l) at (2.75,2.5);
\coordinate (s4-2-u) at (2.75,3);

\draw[init] ($(init-l)-(0,0)$) -- ($(init-u)-(0,0)$) -- ($(init-u)-(0.04,0)$) -- ($(init-l)-(0.04,0)$) -- cycle;

\draw[cone, mainColor] (init-l) -- (s1-1-l) -- (s1-1-u) -- (init-u) -- cycle;
\draw[cone, mainColor] (s1-1-l) -- (s1-1-u) -- (s1-2-u) -- (s1-2-l) -- cycle;
\draw[cone, mainColor] (s2-1-l) -- (s2-1-u) -- (s1-2-u) -- (s2-2-u) -- (s2-2-l) -- cycle;

\draw[border, initloc] (init-l) -- ($(s1-1-l)-(0.0,0)$) -- ($(s1-1-u)-(0.0,0)$) -- (init-u) -- cycle;
\draw[border, path1end] (s2-2-l) -- (s2-1-l) -- (s2-1-u) -- (s1-2-u) -- (s2-2-u) ; %
\draw[border, path1start] (s1-1-l) -- (s1-1-u) -- (s1-2-u) -- (s1-2-l) -- cycle;

\draw[cone, mainColor](s3-2-u) -- (s3-2-l)-- (s3-1-l) -- (s3-1-u);
\draw[cone, mainColor](s4-2-u) -- (s4-2-l)-- (s4-1-l) -- (s4-1-u);

\draw[border,path2start ](s3-2-u) -- (s3-2-l)-- (s3-1-l) -- (s3-1-u); %
\draw[border,path2end ] (s4-2-l)-- (s4-1-l) -- (s4-1-u); %

\fill[refinement] 
($(s2-2-u)-(0.25,0.25)$)-- ($(s1-2-u)+(0.25,0)$) --($(s1-1-u)-(0,0.25)$) -- ($(init-u)-(0.04,0.25)$) 
-- ($(init-u)-(0.04,0)$) -- (s1-1-u) --  (s1-2-u) --  ($(s2-2-u)-(0.25,0.125)$);

\fill[refinement2] 
(goal-u) 
-- ($(s4-1-l)+(0,0.25)$) 
-- (0.5,0.8333) 
-- (init-u)
-- (init-l)
-- (s1-1-l)
-- (1,2.5)
-- (2.5,2.5)
-- cycle;

\draw[border, boxcolor] (goal-l) rectangle (goal-u);
\draw[cone, boxcolor] (goal-l) rectangle (goal-u);

\draw[border, color2] ($(s2-2-u)-(0.25,0.25)$)--($(s2-1-u)+(0.25,0.25)$);
\draw[cone, color2] ($(s2-2-u)-(0.25,0.25)$)--($(s2-1-u)+(0.25,0.25)$) -- (s1-2-u) --  ($(s2-2-u)-(0.25,0.125)$) -- cycle;

\draw[border, color2] ($(s1-2-u)+(0.25,0)$)--($(s1-1-u)-(0,0.25)$);
\draw[cone, color2] ($(s1-2-u)+(0.25,0)$)--($(s1-1-u)-(0,0.25)$) --(s1-1-u) -- (s1-2-u) ;
\draw[border, color2] ($(s1-1-u)-(0,0.25)$)--($(init-u)-(0.04,0.25)$);
\draw[cone, color2] ($(s1-1-u)-(0,0.25)$)--($(init-u)-(0.04,0.25)$) -- ($(init-u)-(0.04,0)$) -- (s1-1-u);

\draw[border, color2] (goal-u) -- ($(s4-1-l)+(0,0.25)$);
\draw[border, color2] ($(s4-1-l)+(0,0.25)$)--($(init-l)+(0.25,0)$);
\draw[border, color2] (0.5,0.8333)--(init-u);

\draw[cone, color2] (goal-u) -- ($(s4-1-l)+(0,0.25)$) -- (s4-1-l) -- ($(goal-u)-(0,0.25)$);
\draw[cone, color2] (s4-1-l)-- ($(s4-1-l)+(0.5,0)$)--($(init-l)+(0.75,0)$)--($(init-l)+(0.25,0)$);
\draw[cone, color2] (0.5,0.8333)--(init-u)--(init-l)--($(init-l)+(0.25,0)$);

\end{tikzpicture}
           \caption{Backward refinement (\tikzboxref)  from goal (\tikzboxgoal) on traces $i=0,1$.}\label{fig:runningex_refinement}%
        \end{subfigure}\hfill%
        \begin{subfigure}[b]{.28\linewidth}
       \centering
           \definecolor{solBase01}{RGB}{88 110 117}
\definecolor{solMagenta}{RGB}{211  54 130}
\definecolor{color2}{named}{solMagenta}
\definecolor{solGreen}{RGB}{133 153   0}
\definecolor{color3}{named}{solGreen}
\definecolor{solYellow}{RGB}{181 137   0}
\definecolor{color4}{named}{solYellow}
\definecolor{dartmouthgreen}{rgb}{0.05, 0.5, 0.06}
\definecolor{color5}{named}{dartmouthgreen}
\definecolor{solBlue}{RGB}{38 139 210}
\definecolor{color6}{named}{solBlue}
\definecolor{davysgrey}{rgb}{0.33, 0.33, 0.33}
\definecolor{fancyblue}{rgb}{0.01, 0.28, 1.0}
\definecolor{amethyst}{rgb}{0.6, 0.4, 0.8}

\definecolor{mainColor}{named}{solBase01}
\definecolor{boxcolor}{named}{color4}
\definecolor{cinitloc}{named}{mainColor}
\definecolor{cpath1light}{named}{color6}
\definecolor{cpath1dark}{named}{fancyblue}
\definecolor{cpath2light}{named}{color3}
\definecolor{cpath2dark}{named}{color5}
\definecolor{crefseg}{named}{color2}
\definecolor{cinterseg}{named}{amethyst}

\tikzstyle{initloc}=[cinitloc]
\tikzstyle{path1start}=[cpath1light,densely dotted]
\tikzstyle{path1end}=[cpath1dark]
\tikzstyle{path2start}=[cpath2light,dotted]
\tikzstyle{path2end}=[cpath2dark]

\tikzstyle{filling}=[very thick, fill, fill opacity = 0.2]
\tikzstyle{cone}=[draw = none, fill, fill opacity = 0.2]
\tikzstyle{border}=[very thick, line join = round]
\tikzstyle{guard}=[fill opacity=0.1]
\tikzstyle{guardticks}=[color3]
\tikzstyle{init}=[border, mainColor, line join = round]
\tikzstyle{refinement}=[pattern=north west lines, pattern color = color2]
\tikzstyle{refinement2}=[pattern=north east lines, pattern color = color2]
\tikzstyle{other}=[opacity = 0.15]%

\tikzstyle{refinedsegment}=[very thick, line join = round, fill, fill opacity = 0.2, color2]
\tikzstyle{intermediatesegment}=[line join = round, fill, cinterseg]

\begin{tikzpicture}[
baseline, remember picture,
font=\footnotesize, scale=1, 
tick/.style={fill=white,font= \scriptsize}]

\useasboundingbox (-0.25,-0.25) rectangle (3,3);

\node (ursprung) at (0,0) {};
\draw[help lines, lightgray, xstep=0.25, ystep=0.25] ($(ursprung)-(0.125,0.125)$) grid (2.875, 3.125);
\node[fill=white] (y) at (0,3) {$y$};
\node[fill=white] (x) at (2.75,0) {$x$};

\foreach \y in {2,4,...,10}  {
    \node[tick] (\y) at (-0.25,0.25*\y) {$\y$}; 
    \draw (-0.05,0.25*\y)--(+0.05,0.25*\y);
}
\foreach \x in {2,4,...,8}  {
    \node[tick] (\x) at (0.25*\x,-0.25) {$\x$}; 
    \draw (0.25*\x,-0.05)--(0.25*\x,+0.05);
}

\draw[-latex,thick] ($(ursprung)-(0,0.125)$) to (y);
\draw[-latex,thick] ($(ursprung)-(0.125,0)$) to (x);

\coordinate (goal-l) at (2,2);
\coordinate (goal-u) at (2.5,2.75);
\coordinate (init-l) at (0.25,0.25);
\coordinate (init-u) at (0.25,0.75);
\coordinate (s1-1-l) at (1,0.25);
\coordinate (s1-1-u) at (1,1);
\coordinate (s1-2-l) at (2.5,1.75);
\coordinate (s1-2-u) at (1.75,1.75);
\coordinate (s2-1-l) at (2,1.25);
\coordinate (s2-1-u) at (1.25,1.25);
\coordinate (s2-2-l) at (2.75,1.625);
\coordinate (s2-2-u) at (2.75,2.25);
\coordinate (s3-1-l) at ($(init-l)$);
\coordinate (s3-1-u) at (0.25,3);
\coordinate (s3-2-l) at ($(s1-1-l)$);
\coordinate (s3-2-u) at (1,3);
\coordinate (s4-1-l) at (0.5,2.5);
\coordinate (s4-1-u) at (0.5,3);
\coordinate (s4-2-l) at (2.75,2.5);
\coordinate (s4-2-u) at (2.75,3);

\draw[init] ($(init-l)-(0,0)$) -- ($(init-u)-(0,0)$) -- ($(init-u)-(0.04,0)$) -- ($(init-l)-(0.04,0)$) -- cycle;

\draw[cone, mainColor] (init-l) -- (s1-1-l) -- (s1-1-u) -- (init-u) -- cycle;
\draw[cone, mainColor] (s1-1-l) -- (s1-1-u) -- (s1-2-u) -- (s1-2-l) -- cycle;
\draw[cone, mainColor] (s2-1-l) -- (s2-1-u) -- (s1-2-u) -- (s2-2-u) -- (s2-2-l) -- cycle;

\draw[border, initloc] (init-l) -- ($(s1-1-l)-(0.0,0)$) -- ($(s1-1-u)-(0.0,0)$) -- (init-u) -- cycle;
\draw[border, path1end] (s2-2-l) -- (s2-1-l) -- (s2-1-u) -- (s1-2-u) -- (s2-2-u) ; %
\draw[border, path1start] (s1-1-l) -- (s1-1-u) -- (s1-2-u) -- (s1-2-l) -- cycle;

\draw[cone, mainColor,other](s3-2-u) -- (s3-2-l)-- (s3-1-l) -- (s3-1-u);
\draw[cone, mainColor,other](s4-2-u) -- (s4-2-l)-- (s4-1-l) -- (s4-1-u);

\draw[border,path2start,other ](s3-2-u) -- (s3-2-l)-- (s3-1-l) -- (s3-1-u); %
\draw[border,path2end,other ] (s4-2-l)-- (s4-1-l) -- (s4-1-u); %

\fill[refinement]  %
($(s2-2-u)-(0.25,0.25)$)-- ($(s1-2-u)+(0.25,0)$) --($(s1-1-u)-(0,0.25)$) -- ($(init-u)-(0.04,0.25)$) 
-- ($(init-u)-(0.04,0)$) -- (s1-1-u) --  (s1-2-u) --  ($(s2-2-u)-(0.25,0.125)$);

\draw[border, boxcolor,other] (goal-l) rectangle (goal-u);
\draw[cone, boxcolor,other] (goal-l) rectangle (goal-u);

\draw[refinedsegment] ($(s2-2-u)-(0.25,0.25)$)--($(s2-1-u)+(0.25,0.25)$) -- (s1-2-u) --  ($(s2-2-u)-(0.25,0.125)$) -- cycle;

\draw[refinedsegment] ($(s1-2-u)+(0.25,0)$)--($(s1-1-u)-(0,0.25)$) --(s1-1-u) -- (s1-2-u) ;
\draw[refinedsegment] ($(s1-1-u)-(0,0.25)$)--($(init-u)-(0.00,0.25)$) -- ($(init-u)-(0.00,0)$) -- (s1-1-u);

\draw[intermediatesegment] ($(s2-2-u)-(0.25,0.25)$)-- (2.25,2) --  ($(s2-2-u)-(0.25,0.125)$) -- cycle;
\draw[intermediatesegment] (2,1.75)--($(s2-1-u)+(0.25,0.25)$) -- (s1-2-u) -- cycle;
\draw[intermediatesegment] (0.99,1) rectangle (1.01,0.75);
\draw[intermediatesegment] (0.24,0.75) rectangle (0.26,0.5);

\node[cpath1dark,font=\scriptsize] at (2.5,1.125) {$\valSet_0^0$};
\node[cpath1light,font=\scriptsize] at (1.875,0.75) {$\valSet_1^0$};
\node[cinitloc,font=\scriptsize] at (1.375,0.2) {$\valSet_2^0$};

\node[cinterseg,font=\scriptsize] at (2.5,2.5) {$\valSet^0_{\phantom{0}}$};
\node[cinterseg,font=\scriptsize] at (1.75,2) {$\JumpSegmentik{0}{1}$};
\node[cinterseg,font=\scriptsize] at (1,1.5) {$\JumpSegmentik{0}{2}$};
\node[cinterseg,font=\scriptsize] at (0.25,1) {$\JumpSegmentik{0}{3}$};

\node[crefseg,font=\scriptsize] at (2.125,2.25) {$\RefinedSegmentik{0}{0}$};
\node[crefseg,font=\scriptsize] at (1.375,1.75) {$\RefinedSegmentik{0}{1}$};
\node[crefseg,font=\scriptsize] at (0.625,1.25) {$\RefinedSegmentik{0}{2}$};

\end{tikzpicture}
           \caption{Alternating segments $\JumpSegmentik{0}{k}$ and $\RefinedSegmentik{0}{k}$ on trace $i=0$.%
           }\label{fig:runningex_refinement_names}%
        \end{subfigure}\hfill%
        \begin{subfigure}[b]{.3\linewidth}
       \centering
           \begin{tikzpicture}[
scale=1,
node distance = 0.4cm and 0.2cm,
baseline,
remember picture,
n/.style={draw, text width = 0.8cm, text = black, %
align = center, font= \footnotesize, rounded corners, very thick, execute at begin node=\setlength{\baselineskip}{8pt}%
},
en/.style={draw=none, minimum height=0cm, font = \scriptsize, align = center},%
ref/.style={draw, circle, minimum width=1.5mm, inner sep=0, fill,refinementcolor},%
c/.style={draw, fill, black, circle, inner sep=0, outer sep=0, minimum size=1mm},
l/.style={anchor=west, inner sep=0, font=\footnotesize},
]

\useasboundingbox (-2,0.25) rectangle (2,-3);

\node[n, draw=cinitloc](l0) at (0,0) {	$\ell_0$	};
\node[n, draw=cpath1light,below left = of l0.south]  (l1)  {$\ell_1$};
\node[n, draw=cpath1dark, below = of l1] (l2)  {$\ell_2$};
\node[n, draw=cpath2light, below right = of l0.south] (l3) {$\ell_3$};
\node[n, draw=cpath2dark, below = of l3] (l4) {$\ell_4$ };

\node[n, draw=none, below = 2mm of l2] (p0ph) {$\States^0$};
\node[n, draw=none, below = 2mm of l4] (p1ph) {$\States^1$};

 \draw[-latex, very thick] (l0) to node[en, left] {}  (l1);
 \draw[-latex, very thick] (l0) to node[en, above] {} (l3);
 \draw[-latex, very thick] (l1) to node[en, left] {} (l2);
 \draw[-latex, very thick] (l3) to node[en, left] {} (l4);

\node[ref, left = 2mm of l0.center,] (p0l0) {};
\node[ref, left = 2mm of l1.center,] (p0l1) {};
\node[ref, left = 2mm of l2.center,] (p0l2) {};
\node[ref, left = 2mm of p0ph.center,] (p0) {};

\node[ref, right = 2mm of l0.center,] (p1l0) {};
\node[ref, right = 2mm of l3.center,] (p1l3) {};
\node[ref, right = 2mm of l4.center,] (p1l4) {};
\node[ref, right = 2mm of p1ph.center,] (p1) {};

\node[n, draw=none, below = 1mm of p0,font = \scriptsize] (i0) {$i=0$};
\node[n, draw=none, below = 1mm of p1,font = \scriptsize] (i1) {$i=1$};

\draw[refinementcolor,thick] (p0l0) -- (p0l1) -- (p0l2) -- (p0);
\draw[refinementcolor,thick] (p1l0) -- (p1l3) -- (p1l4) -- (p1);

\end{tikzpicture}%
           \caption{Reach tree with traces $i=0,1$ leading to $\Goali$ via refined segments $\RefinedSegmentik{i}{k}$, $\valSet^i$.}\label{fig:targettree}%
        \end{subfigure}
    \caption{Backward refinement.}
    \label{fig:backward_refinement}
\end{figure}
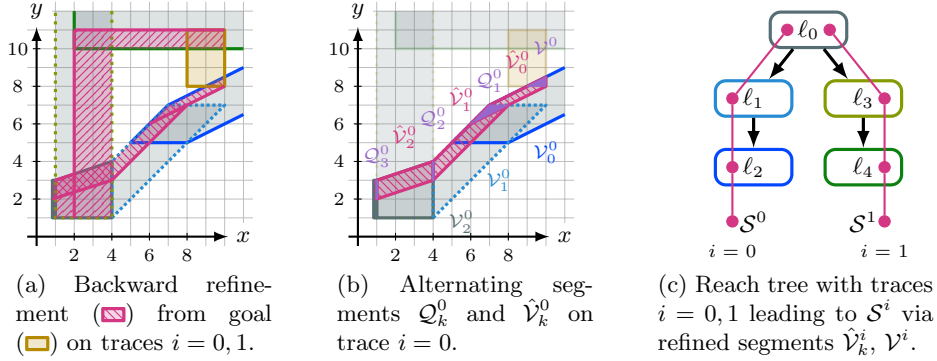
\begin{definition}
Given an intermediate goal segment $\JumpSegmentik{i}{k}$ within a segment $\Goalvalik{k}$ and its corresponding location $\ell$, the \emph{refined segment} $\Refvalik{k}$ on trace $i$ is 
    $\Refvalik{k} = \btc{\ell}( \JumpSegmentik{i}{k} ) \cap \Goalvalik{k}$.
The first intermediate goal segment $\JumpSegmentik{i}{0}$ on trace $i$ is $\Goalvali$ itself. 
\end{definition}
Figure~\ref{fig:backward_refinement_steps}  illustrates refined segments computed from intermediate goal segments, corresponding to the running example.
Given a transition $e$ from $\ell_p$ to $\ell$ connecting segments $\Goalvalik{k+1}$ and $\Goalvalik{k}$, we compute the backward one-step relation of a refined segment $\Refvalik{k}$.
By intersecting with the next %
forward segment $\Goalvalik{k+1}$ it is ensured, that the resulting segment $\JumpSegmentik{i}{k+1}$ is a subset of $\Goalvalik{k+1}$.
This will be used as intermediate goal segment for the next computation step.

\begin{definition}
For a refined segment $\Refvalik{k}$ and its corresponding location $\ell$, the  next \emph{intermediate goal segment} $\JumpSegmentik{i}{k+1}$ 
on trace $i$ is 
    $\JumpSegmentik{i}{k+1} =\bosr{e}(\Refvalik{k}) \cap \Goalvalik{k+1}$.
If $\Refvalik{k}$ corresponds to an initial location $\ell_0$ (i.e., $\Init(\ell_0)\not=\emptyset$), 
we use the initial set of $\ell_0$ instead of $\Goalvalik{k+1}$ for the intersection, i.e., $\JumpSegmentik{i}{k+1} =\bosr{e}(\Refvalik{k}) \cap \Init(\ell_0)$.
\end{definition}

\begin{runningexample} 
Backward refinement  for the running example is illustrated in Figure~\ref{fig:runningex_refinement} and Figure~\ref{fig:runningex_refinement_names}. 
Figure~\ref{fig:targettree} illustrates the two traces $i=0,1$. %
Starting from segments $\valSet^0$ and $\valSet^1$, the refined segments $\hat{\valSet}^0_k$ and $\hat{\valSet}^1_k$ can be computed iteratively. 
Figure~\ref{fig:backward_refinement_steps} illustrates one refined segment on each trace. 
The refined segment $\RefinedSegmentik{0}{2}$ and the next intermediate goal segment $\JumpSegmentik{0}{3}$  (c.f. Figure~\ref{fig:backward_ref_steps0}) can be computed from $\JumpSegmentik{0}{2}$.
In addition to the constraints visible in Figure~\ref{fig:backward_ref_steps0}, $\RefinedSegmentik{0}{2}$ then contains the constraint $\val_r = \val_x-1$, stemming from the initial valuation of $\val_x=1$.
Note that $\JumpSegmentik{0}{3}$ is computed via intersection with $\Init(\ell_0)$.

The refined segment $\RefinedSegmentik{1}{1}$ and the next intermediate goal segment $\JumpSegmentik{1}{2}$ (c.f. Figure~\ref{fig:backward_ref_steps1}) are computed from $\JumpSegmentik{1}{1}$.
Again, $\RefinedSegmentik{1}{1}$ and $\JumpSegmentik{1}{2}$ contain the constraint $\val_r = \val_x-1$.
Note that, since also $\val_x\in [2,4]$, $\val_r$ is implicitely restricted to be contained by $[1,3]$. 
These computation steps can also be found in~\ref{appendix:runningexample}.%
\end{runningexample}
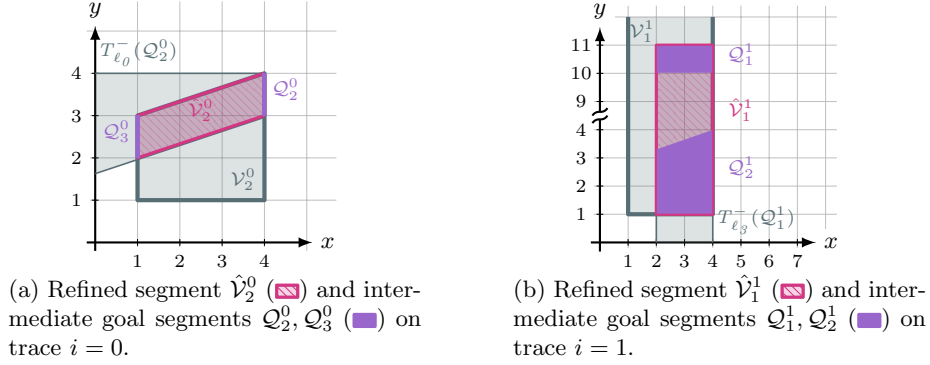
\begin{figure}[tb]
    \centering
       \begin{subfigure}[b]{.45\linewidth}
       \centering
           \definecolor{solBase01}{RGB}{88 110 117}
\definecolor{solMagenta}{RGB}{211  54 130}
\definecolor{color2}{named}{solMagenta}
\definecolor{solGreen}{RGB}{133 153   0}
\definecolor{color3}{named}{solGreen}
\definecolor{solYellow}{RGB}{181 137   0}
\definecolor{color4}{named}{solYellow}
\definecolor{dartmouthgreen}{rgb}{0.05, 0.5, 0.06}
\definecolor{color5}{named}{dartmouthgreen}
\definecolor{solBlue}{RGB}{38 139 210}
\definecolor{color6}{named}{solBlue}
\definecolor{davysgrey}{rgb}{0.33, 0.33, 0.33}
\definecolor{fancyblue}{rgb}{0.01, 0.28, 1.0}
\definecolor{amethyst}{rgb}{0.6, 0.4, 0.8}

\definecolor{mainColor}{named}{solBase01}
\definecolor{boxcolor}{named}{color4}
\definecolor{cinitloc}{named}{mainColor}
\definecolor{cpath1light}{named}{color6}
\definecolor{cpath1dark}{named}{fancyblue}
\definecolor{cpath2light}{named}{color3}
\definecolor{cpath2dark}{named}{color5}
\definecolor{crefseg}{named}{color2}
\definecolor{cinterseg}{named}{amethyst}

\tikzstyle{initloc}=[cinitloc]
\tikzstyle{path1start}=[cpath1light,densely dotted]
\tikzstyle{path1end}=[cpath1dark]
\tikzstyle{path2start}=[cpath2light,dotted]
\tikzstyle{path2end}=[cpath2dark]

\tikzstyle{filling}=[very thick, fill, fill opacity = 0.2]
\tikzstyle{cone}=[draw = none, fill, fill opacity = 0.2]
\tikzstyle{border}=[very thick, line join = round]
\tikzstyle{guard}=[fill opacity=0.1]
\tikzstyle{guardticks}=[color3]
\tikzstyle{init}=[border, mainColor, line join = round]
\tikzstyle{refinement}=[pattern=north west lines, pattern color = color2]
\tikzstyle{refinement2}=[pattern=north east lines, pattern color = color2]
\tikzstyle{other}=[opacity = 0.15]%

\tikzstyle{refinedsegment}=[very thick, line join = round, fill, fill opacity = 0.2, color2]
\tikzstyle{intermediatesegment}=[line join = round, fill, cinterseg]

\begin{tikzpicture}[
baseline, remember picture,
font=\footnotesize, scale=2.25, 
tick/.style={fill=white,font= \scriptsize}]

\useasboundingbox (-0.11,-0.11) rectangle (1.48,1.48);

\node (ursprung) at (0,0) {};
\draw[help lines, lightgray, xstep=0.25, ystep=0.25] ($(ursprung)-(0.05,0.05)$) grid (1.425, 1.425);
\node[fill=white] (y) at (0,1.375) {$y$};
\node[fill=white] (x) at (1.375,0) {$x$};
\draw[-latex,thick] ($(ursprung)-(0,0.05)$) to (y);
\draw[-latex,thick] ($(ursprung)-(0.05,0)$) to (x);

\node[tick] (1) at (0.25,-0.11) {$1$};
\node[tick] (2) at (0.5,-0.11) {$2$};
\node[tick] (3) at (0.75,-0.11) {$3$};
\node[tick] (4) at (1,-0.11) {$4$};
\draw ($(0.25,0)+(0,0.02)$) -- ($(0.25,0)-(0,0.02)$);
\draw ($(0.5,0)+(0,0.02)$) -- ($(0.5,0)-(0,0.02)$);
\draw ($(0.75,0)+(0,0.02)$) -- ($(0.75,0)-(0,0.02)$);
\draw ($(1,0)+(0,0.02)$) -- ($(1,0)-(0,0.02)$);

\node[tick] (1) at (-0.11,0.25) {$1$};
\node[tick] (2) at (-0.11,0.5) {$2$};
\node[tick] (3) at (-0.11,0.75) {$3$};
\node[tick] (4) at (-0.11,1) {$4$};
\draw ($(0,0.25)+(0.02,0)$) -- ($(0,0.25)-(0.02,0)$);
\draw ($(0,0.5)+(0.02,0)$) -- ($(0,0.5)-(0.02,0)$);
\draw ($(0,0.75)+(0.02,0)$) -- ($(0,0.75)-(0.02,0)$);
\draw ($(0,1)+(0.02,0)$) -- ($(0,1)-(0.02,0)$);

\coordinate (goal-l) at (2,2);
\coordinate (goal-u) at (2.5,2.75);
\coordinate (init-l) at (0.25,0.25);
\coordinate (init-u) at (0.25,0.75);
\coordinate (s1-1-l) at (1,0.25);
\coordinate (s1-1-u) at (1,1);
\coordinate (s1-2-l) at (2.5,1.75);
\coordinate (s1-2-u) at (1.75,1.75);
\coordinate (s2-1-l) at (2,1.25);
\coordinate (s2-1-u) at (1.25,1.25);
\coordinate (s2-2-l) at (2.75,1.625);
\coordinate (s2-2-u) at (2.75,2.25);
\coordinate (s3-1-l) at ($(init-l)$);
\coordinate (s3-1-u) at (0.25,3);
\coordinate (s3-2-l) at ($(s1-1-l)$);
\coordinate (s3-2-u) at (1,3);
\coordinate (s4-1-l) at (0.5,2.5);
\coordinate (s4-1-u) at (0.5,3);
\coordinate (s4-2-l) at (2.75,2.5);
\coordinate (s4-2-u) at (2.75,3);

\draw[cone, mainColor] (init-l) -- (s1-1-l) -- (s1-1-u) -- (init-u) -- cycle;

\draw[border, initloc,ultra thick] (init-l) -- ($(s1-1-l)-(0.0,0)$) -- ($(s1-1-u)-(0.0,0)$) -- (init-u) -- cycle;

\draw[border,initloc, semithick] (0,1) -- (1,1) -- (1,0.74) -- (0.25,0.49)--(0,0.406666);
\draw[cone,initloc] (0,1) -- (1,1) -- (1,0.74) -- (0.25,0.49)--(0,0.406666);

\draw[refinedsegment] ($(s1-1-u)-(0,0.25)$)--($(init-u)-(0.00,0.25)$) -- ($(init-u)-(0.00,0)$) -- (s1-1-u);
\fill[refinement,opacity=0.5] ($(s1-1-u)-(0,0.25)$)--($(init-u)-(0.00,0.25)$) -- ($(init-u)-(0.00,0)$) -- (s1-1-u);

\draw[intermediatesegment] (0.99,1) rectangle (1.01,0.75);
\draw[intermediatesegment] (0.24,0.75) rectangle (0.26,0.5);

\node[cinitloc,font=\scriptsize] at (0.875,0.375) {$\valSet_2^0$};

\node[cinterseg,font=\scriptsize] at (1.125,0.875) {$\JumpSegmentik{0}{2}$};
\node[cinterseg,font=\scriptsize] at (0.125,0.625) {$\JumpSegmentik{0}{3}$};

\node[crefseg,font=\scriptsize] at (0.625,0.75) {$\RefinedSegmentik{0}{2}$};

\node[initloc,font=\scriptsize] at (0.25,1.125) {$\btc{\ell_0}(\JumpSegmentik{0}{2})$};

\end{tikzpicture}
            \caption{Refined segment $\RefinedSegmentik{0}{2}$ (\tikzboxref) and intermediate goal segments $\JumpSegmentik{0}{2},\JumpSegmentik{0}{3}$  (\tikzboxintermediateseg) on trace $i=0$.}\label{fig:backward_ref_steps0}
        \end{subfigure}\hfill%
        \begin{subfigure}[b]{.45\linewidth}
        \centering
           \definecolor{solBase01}{RGB}{88 110 117}
\definecolor{solMagenta}{RGB}{211  54 130}
\definecolor{color2}{named}{solMagenta}
\definecolor{solGreen}{RGB}{133 153   0}
\definecolor{color3}{named}{solGreen}
\definecolor{solYellow}{RGB}{181 137   0}
\definecolor{color4}{named}{solYellow}
\definecolor{dartmouthgreen}{rgb}{0.05, 0.5, 0.06}
\definecolor{color5}{named}{dartmouthgreen}
\definecolor{solBlue}{RGB}{38 139 210}
\definecolor{color6}{named}{solBlue}
\definecolor{davysgrey}{rgb}{0.33, 0.33, 0.33}
\definecolor{fancyblue}{rgb}{0.01, 0.28, 1.0}
\definecolor{amethyst}{rgb}{0.6, 0.4, 0.8}

\definecolor{mainColor}{named}{solBase01}
\definecolor{boxcolor}{named}{color4}
\definecolor{cinitloc}{named}{mainColor}
\definecolor{cpath1light}{named}{color6}
\definecolor{cpath1dark}{named}{fancyblue}
\definecolor{cpath2light}{named}{color3}
\definecolor{cpath2dark}{named}{color5}
\definecolor{crefseg}{named}{color2}
\definecolor{cinterseg}{named}{amethyst}

\tikzstyle{initloc}=[cinitloc]
\tikzstyle{path1start}=[cpath1light,densely dotted]
\tikzstyle{path1end}=[cpath1dark]
\tikzstyle{path2start}=[cpath2light,dotted]
\tikzstyle{path2end}=[cpath2dark]

\tikzstyle{filling}=[very thick, fill, fill opacity = 0.2]
\tikzstyle{cone}=[draw = none, fill, fill opacity = 0.2]
\tikzstyle{border}=[very thick, line join = round]
\tikzstyle{guard}=[fill opacity=0.1]
\tikzstyle{guardticks}=[color3]
\tikzstyle{init}=[border, mainColor, line join = round]
\tikzstyle{refinement}=[pattern=north west lines, pattern color = color2]
\tikzstyle{refinement2}=[pattern=north east lines, pattern color = color2]
\tikzstyle{other}=[opacity = 0.15]%

\tikzstyle{refinedsegment}=[very thick, line join = round, fill, fill opacity = 0.2, color2]
\tikzstyle{intermediatesegment}=[line join = round, fill, cinterseg]

\tikzset{ext/.pic={
\path [fill=white,scale=0.5] (-0.2,0)to[bend left](0,0.1)to[bend right](0.2,0.2)to(0.2,0)to[bend left](0,-0.1)to[bend right](-0.2,-0.2)--cycle;
\draw[scale=0.5] (-0.2,0)to[bend left](0,0.1)to[bend right](0.2,0.2) (0.2,0)to[bend left](0,-0.1)to[bend right](-0.2,-0.2);
}}

\begin{tikzpicture}[
baseline, remember picture,
font=\footnotesize, scale=1.5, 
tick/.style={fill=white,font= \scriptsize,inner sep= 1mm}]

\useasboundingbox (-0.166,-0.166) rectangle (2.22,2.22);

\node (ursprung) at (0,0) {};
\draw[help lines, lightgray, xstep=0.25, ystep=0.25] ($(ursprung)-(0.08,0.08)$) grid (2.1,2.1);
\node[fill=white] (y) at (0,2.05) {$y$};
\node[fill=white] (x) at (2.05,0) {$x$};
\draw[thick] ($(ursprung)-(0,0.08)$) to  (0,1);
\draw[-latex,thick] ($(ursprung)-(0.08,0)$) to (x);

\node[tick] (1) at (0.25,-0.166) {$1$};
\node[tick] (2) at (0.5,-0.166) {$2$};
\node[tick] (3) at (0.75,-0.166) {$3$};
\node[tick] (4) at (1,-0.166) {$4$};
\node[tick] (1) at (1.25,-0.166) {$5$};
\node[tick] (2) at (1.5,-0.166) {$6$};
\node[tick] (3) at (1.75,-0.166) {$7$};
\draw ($(0.25,0)+(0,0.03)$) -- ($(0.25,0)-(0,0.03)$);
\draw ($(0.5,0)+(0,0.03)$) -- ($(0.5,0)-(0,0.03)$);
\draw ($(0.75,0)+(0,0.03)$) -- ($(0.75,0)-(0,0.03)$);
\draw ($(1,0)+(0,0.03)$) -- ($(1,0)-(0,0.03)$);
\draw ($(1.25,0)+(0,0.03)$) -- ($(1.25,0)-(0,0.03)$);
\draw ($(1.5,0)+(0,0.03)$) -- ($(1.5,0)-(0,0.03)$);
\draw ($(1.75,0)+(0,0.03)$) -- ($(1.75,0)-(0,0.03)$);

\node[tick] (1) at (-0.166,0.25) {$1$};
\node[tick] (2) at (-0.166,0.5) {$2$};
\node[tick] (3) at (-0.166,0.75) {$3$};
\node[tick] (4) at (-0.166,1) {$4$};
\node[tick] (9) at (-0.166,1.25) {$9$};
\node[tick] (10) at (-0.166,1.5) {$10$};
\node[tick] (11) at (-0.166,1.75) {$11$};
\draw ($(0,0.25)+(0.03,0)$) -- ($(0,0.25)-(0.03,0)$);
\draw ($(0,0.5)+(0.03,0)$) -- ($(0,0.5)-(0.03,0)$);
\draw ($(0,0.75)+(0.03,0)$) -- ($(0,0.75)-(0.03,0)$);
\draw ($(0,1)+(0.03,0)$) -- ($(0,1)-(0.03,0)$);
\draw ($(0,1.25)+(0.03,0)$) -- ($(0,1.25)-(0.03,0)$);
\draw ($(0,1.5)+(0.03,0)$) -- ($(0,1.5)-(0.03,0)$);
\draw ($(0,1.75)+(0.03,0)$) -- ($(0,1.75)-(0.03,0)$);

\draw[thick] (0,1) --pic{ext} (0,1.25);
\draw[-latex,thick] (0,1.25) to  (y);

\coordinate (goal-l) at (2,2);
\coordinate (goal-u) at (2.5,2.75);
\coordinate (init-l) at (0.25,0.25);
\coordinate (init-u) at (0.25,0.75);
\coordinate (s1-1-l) at (1,0.25);
\coordinate (s1-1-u) at (1,1);
\coordinate (s1-2-l) at (2.5,1.75);
\coordinate (s1-2-u) at (1.75,1.75);
\coordinate (s2-1-l) at (2,1.25);
\coordinate (s2-1-u) at (1.25,1.25);
\coordinate (s2-2-l) at (2.75,1.625);
\coordinate (s2-2-u) at (2.75,2.25);
\coordinate (s3-1-l) at ($(init-l)$);
\coordinate (s3-1-u) at (0.25,3);
\coordinate (s3-2-l) at ($(s1-1-l)$);
\coordinate (s3-2-u) at (1,3);
\coordinate (s4-1-l) at (0.5,2.5);
\coordinate (s4-1-u) at (0.5,3);
\coordinate (s4-2-l) at (2.75,2.5);
\coordinate (s4-2-u) at (2.75,3);

\draw[border, initloc,ultra thick]  (0.25,2)--(0.25,0.25) --(1,0.25)--(1,2);
\draw[cone, initloc,ultra thick]  (0.25,2)--(0.25,0.25) --(1,0.25)--(1,2);

\draw[border,initloc, semithick] (1,0)--(1,1.75) --  (0.5,1.75)--(0.5,0);
\draw[cone,initloc]  (1,0)--(1,1.75) --  (0.5,1.75)--(0.5,0);

\draw[refinedsegment] (1,1.75) --  (0.5,1.75)--(0.5,0.25) --(1,0.25) --cycle;
\fill[refinement,opacity=0.5] (1,1.75) --  (0.5,1.75)--(0.5,0.25) --(1,0.25) --cycle;

\draw[intermediatesegment] (0.51,1.51) rectangle (0.99,1.74);
\draw[intermediatesegment] (0.99,0.99) -- (0.51,0.823) -- (0.51,0.26) -- (0.99,0.26) -- cycle;

\node[cinitloc,font=\scriptsize] at (0.375,1.875) {$\valSet_1^1$};

\node[cinterseg,font=\scriptsize] at (1.25,0.62) {$\JumpSegmentik{1}{2}$};
\node[cinterseg,font=\scriptsize] at (1.25,1.62) {$\JumpSegmentik{1}{1}$};

\node[crefseg,font=\scriptsize] at (1.25,1.125) {$\RefinedSegmentik{1}{1}$};

\node[initloc,font=\scriptsize] at (1.375,0.2) {$\btc{\ell_3}(\JumpSegmentik{1}{1})$};

\end{tikzpicture}
           \caption{Refined segment $\RefinedSegmentik{1}{1}$ (\tikzboxref) and intermediate goal segments $\JumpSegmentik{1}{1},\JumpSegmentik{1}{2}$  (\tikzboxintermediateseg) on trace $i=1$.}\label{fig:backward_ref_steps1}
        \end{subfigure}
    \caption{Detailed backward refinement steps.}
    \label{fig:backward_refinement_steps}
\end{figure}

\subsection{Extracting the sample domain}\label{subsec:collection}

Backward refinement  results in refined segments $\Refvalik{k}$, from which the sample domain for all random variables is extracted as follows.
Refined segments contain information on interdependencies between all variables and valuations leading to the goal in $\Goalvali$.
The sample domain for each random variable $\randomvarij$ is derived from the segment which corresponds to the $n$-th expiration of random clock $\clocki$.

Polytopes $\Polymaxi\subset \Reals^{\dimR}$ collect the sample values which lead to the goal via trace $i$.
Traversing the traces in a forward manner, for each segment we derive information on the sample domains from the constraints on the valuations of the random clocks which allow taking the next step on trace $i$ (in the reach tree). %
Thus, we collect all valid values for each random delay on trace $i$.

For each segment,  constraints on the samples are derived as follows.
First, we  project the segment onto the stochastic dimensions.
Second, we collect information about all random delays, which are either  already expired,  about to expire in the next step on the trace, or not expired.
(i) Random delays that are already expired cannot provide any new information, and hence do not have to be considered again.
(ii) A step in the trace that corresponds to the expiration of a random delay, induces (upper) bounds on the sample domain.
Each edge in the reach tree maps to exactly one jump in the automaton. 
In case of a stochastic jump, the step results in the upper bounds of exactly one random clock. 
(iii) To account for information on future expirations, the upper bounds of all other random delays have to be \emph{lifted}. %

\begin{definition}
    We define the \emph{lifting} for a variable $\clocki$ in a polytope $P \subseteq \Reals^\dimR$ as $\Lift{\clocki}{P}{:=\{(\sample_1,\dots,\smash{\sample_r+ c},\dots,\sample_{\dimR}) \in \Reals^\dimR  \!\mid\! (\sample_1,\dots,\sample_r,\dots,\sample_{\dimR})  \in P \!\land c \in [0,\infty) \}}$.
\end{definition}%

This iterative collection of constraints on the sample domain leads to a polytope $\Polymaxi$ that contains all sample values which allow to follow trace $i$. %
To compute maximum reachability probabilities, 
all possibilities of reaching the goal have to be included in the integration.
This corresponds to taking the union over all $\Polymaxi$ for $i \in \pathindices$.

Summarizing, this leads to a polytope $\Polymax= \bigcup_{i \in \pathindices} \Polymaxi $.

\begin{runningexample}
The random delay $r$ does not expire on trace $0$, hence  the lower constraints on the sample domain for $\sample_r$ are iteratively collected, leading to the strongest lower constraint  in $\valSet^0$.
By projecting $\valSet^0$ onto $\Reals^\dimR$ and lifting the resulting polytope in dimension $r$, the constraint $\sample_r\geq 3$ can be derived.

    On trace $1$, the expiration of the random clock corresponds to  the transition from $\ell_0$ to $\ell_3$, i.e., $r$ is about to expire in segment $\RefinedSegmentik{1}{2}$.
    Hence, from $\RefinedSegmentik{1}{2}$, the constraints $\sample_r \geq 1$ and $\sample_r \leq 3$ are derived, i.e. $\Polymaxiparam{1}=[1,3]$.
    This leads to $\Polymax=[3,\infty) \cup [1,3] = [1,\infty)$, 
which contains all values for random variable $\randomvar_r$ for which the goal is reachable: for $\sample_r\geq 3$ via trace $0$, and for $\sample_r\in [1,3]$ via trace $1$. %
\end{runningexample}

\subsection{Maximum prophetic reachability probabilities}\label{subsec:schedulers}
To maximize continuous nondeterminism prophetically, we  partition  the potentially infinite set of prophetic schedulers $\SchedulersProphMax$ with respect to their ability to reach the goal.
The backward refinement returns the fragment of the reach tree that leads to goal states $\Goal$ (and specifically state sets $\Goali \in \Goalreached$), that allows extracting the sample domain.
This process incorporates knowledge of future expiration times of random variables, leading to prophetic schedulers.

Reachability  defines an equivalence relation on the set of schedulers $\SchedulersProphMax$ with respect to the  state sets $\Goali$, reachable via trace $i$.
This results in equivalence classes $\SchedulersRepr$, which contain all schedulers $\scheduler$ able to reach $\Goali$. Hence, $\bigcup_{i\in \pathindices} \SchedulersRepr = \SchedulersProphMax$.
Via this equivalence relation, we are 
able to resolve different types of (continuous) nondeterminism.
This is explained in the following for nondeterministic time delays on transitions, rectangular flow sets and conflicting transitions.

  \paragraph{Initial and time nondeterminism.} 
    Taking a transition (from $\ell_p$ to $\ell$) at different points in time  leads to different states in the set, with which target location $\ell$ is entered.
    This set $\JumpSegmentik{i}{k+1} \subset \RefinedSegmentik{i}{k}$
    then contains all states corresponding to time delays which enable reaching 
    $\JumpSegmentik{i}{k}$.
    Recall, that at the end of the trace, $\JumpSegmentik{i}{0}$ equals $\Goalvali$.
    This corresponds to a maximizing scheduler choosing (in a forward way) a time delay for that transition, such that from each state in $\JumpSegmentik{i}{k+1}$, entering location $\ell_p$, it is possible to reach the intermediate goal segment $\JumpSegmentik{i}{k}$.
 Similarly, for a corresponding initial location, $\JumpSegmentik{i}{k+1}$ restricts the initial set.
The scheduler then chooses an initial state, such that  $\JumpSegmentik{i}{k}$ can be reached via $\RefinedSegmentik{i}{k}$.
 
     Figure~\ref{fig:backward_ref_steps1} illustrates overlapping segments caused by a nondeterministic guard.
    The backward time closure from $\JumpSegmentik{1}{1}$ in
    segment $\valSet^1_1$
    restricts the range of possible time delays: %
    the (intermediate) goal segment can solely be reached  from the pink fragment, 
    i.e., from states with $\val_y\leq11$.
    
\paragraph{Rate nondeterminism.} 
    The backward refinement results in restricted segments, which implicitly define a partitioning of the schedulers. 
    For every state in $\JumpSegmentik{i}{k+1}$, %
    a maximizing prophetic scheduler can pick at least  one slope, which leads to the intermediate goal set $\JumpSegmentik{i}{k}$. 
    In contrast, for all states outside of $\JumpSegmentik{i}{k+1}$, such a slope does not exist. 
  Figure~\ref{fig:backward_ref_steps0} illustrates that only initial states in $\RefinedSegmentik{0}{2}$ (and hence in $\JumpSegmentik{0}{3}$) can reach the goal. 
  The choice of the initial state restricts the possible rates.
     In this example, for initial state $\val_x=1, \val_y=2$, only the largest possible rate ($\dot{y}=\nicefrac{1}{3}$) leads to the intermediate goal segment $\JumpSegmentik{0}{2}$ and enables reaching $\valSet^0$.

\paragraph{Discrete nondeterminism.} Every discrete choice leads to a different trace in the forward flowpipe. The backward analysis starts from $\Goalreached$, \ie only considers  traces leading to goal states. 
Hence, the union of all $\Polymaxi$ %
over all traces $i$, obtained from backward refinement, represents all discrete choices which reach the goal. 
Consequently, this maximizes discrete nondeterminism.

Summarizing, the valuations induced by all maximizing schedulers match the valuation sets computed by the backward refinement. 
\begin{lemma}\label{lemma:union}
Given the set of valuations $\valSet^\scheduler$ over the space of the random variables that allow scheduler $\scheduler$ to reach the goal set $\Goal$,
and an equivalence relation over $\SchedulersProphMax$ defining equivalence classes $\SchedulersRepr$, 
$\bigcup_{\scheduler\in\SchedulersProphMax}\valSet^\scheduler$ can be computed as follows:
\begin{align} \label{eq:integrationdomain}
    \bigcup_{\scheduler\in\SchedulersProphMax}\valSet^\scheduler 
    &= \bigcup_{i \in \pathindices}\bigcup_{\scheduler \in \SchedulersRepr}\valSet^\scheduler
    = \bigcup_{i \in \pathindices} \valSet ^ \SchedulersRepr
    = \bigcup_{i \in \pathindices} \Polymaxi
    =\Polymax,
\end{align}
 where
 $\valSet^\SchedulersRepr$ is the %
 union of valuations induced by  schedulers $\scheduler\in \SchedulersRepr$.
\end{lemma}

For a proof of Lemma~\ref{lemma:union}, we refer to~\ref{appendix:proofs}. %
We can now compute the  maximum reachability probability by integration over $\Polymax$:
 \begin{equation}\label{eq:probability}
p^{\SchedulersProphetic}_\textsl{max}(\Goal, \tmax) 
= \int_{\bigcup_{\scheduler \in \SchedulersProphMax}\valSet^\scheduler} G(\assignment) \ d\assignment
= \int_{\Polymax} G(\assignment) d\assignment
,
\end{equation}
where  $G(\assignment)= \prod_{\contrandomvar\in \VarRandom} \Distr(\contrandomvar) $ is the joint probability density function. %

The maximum reachability probability stems from all schedulers which can reach the goal. Hence, taking the union over all sample values $\valSet^\scheduler$ reached by  $\scheduler\in\SchedulersProphMax$ results in the integration domain for maximum reachability probabilities.
By construction, $\SchedulersProphMax$ contains all maximizing prophetic schedulers:
\begin{compactenum}[(i)]
\item All schedulers $\scheduler$ able to reach the goal $\Goal$, reach a state set $\Goali$ and are hence represented by an equivalence class $\SchedulersRepr$, collectively reaching $\Goali$. 
\item Schedulers $\scheduler\in \SchedulersProphetic\setminus\SchedulersProphMax$ cannot reach the goal states $\Goal$. %
In case a scheduler $\scheduler \in \SchedulersProphetic\setminus\SchedulersProphMax$ could reach a goal state,  the state(s) reachable by $\scheduler$ would belong to a  flowpipe segment, and hence belong to a state set $\Goali$.
\item  The combination of forward analysis and backward refinement results in prophetic schedulers $\SchedulersProphMax$. 
The former ensures that all  sample values leading to goal states are known. 
The latter partitions the infinite set of schedulers  w.r.t the reachability of different state sets $\Goali$ using this knowledge.
 
\end{compactenum}

\paragraph{Integration error.}
Multidimensional integration over unbounded polytopes is in practice not possible. Hence, we lift to a predefined \emph{integration bound} $\tint \geq \tmax$  and not to infinity, as stated in Section~\ref{subsec:collection}. %
This results in an overapproximation error:
\begin{align*}
   e_{\infty}= 1 -  \int_{[0,\tint]^{\dimR}} G(\assignment) d\assignment.
\end{align*}
This error is exact if: (i) no random clock has ever expired upon reaching the goal set on all traces, or (ii) the support of all random clocks that have not expired upon reaching the goal is finite.
Clearly, increasing $\tint$ decreases $e_{\infty}$.
Integration is done statistically with Monte Carlo Vegas~\cite{Lepage1978MonteCarloVegas},  which introduces an additional statistical error $\estat$,  depending on the number of integration samples.

\begin{runningexample}
Maximizing schedulers  form two equivalence classes, where $\SchedulersRepri{0}$ contains all schedulers reaching $\States^0$ and $\SchedulersRepri{1}$ contains all schedulers reaching $\States^1$.
Schedulers in $\SchedulersRepri{0}$ start  with  $\val_y \geq 2$ in $\ell_0$, 
then pick a rate of $y$ that is at least $1-\nicefrac{\val_y}{3}$ in $\ell_1$. 
Finally, the time delay in $\ell_1$ has to ensure $\val_y \geq 3 + \nicefrac{1}{2}\cdot \val_x$.
 Schedulers in $\SchedulersRepri{1}$ choose to leave  $\ell_3$ such that $\val_y \leq 11$. 

Assuming a folded normal distribution $\mathcal{N}(2,1)$ for the stochastic delay,  the maximum reachability probability to reach $\Goal$ before $\tmax = 10$ is computed by integration over $[1,\infty]$:
$p^{\SchedulersProphetic}_\textsl{max}(\Goal, 10) 
= \int_{[1,\infty]} G(\assignment) d\assignment
= 1- 0.2464
= 0.7536
.$
\end{runningexample}

\paragraph{Computational complexity.}\label{subsec:complexity}
The complexity of forward reachability analysis $\mathit{fwra}$ is exponential in the state-space dimension and depends on the automaton, the set of initial states, and the number of jumps $\jumpmax$ depending on $\tmax$. 
The complexity of computing one refined segment $\RefinedSegmentik{i}{k}$ from its predecessor $\RefinedSegmentik{i}{k-1}$ 
is denoted $\mathit{bwra}$  
with worst-case length $\mathcal{O}(\jumpmax)$ of traces $i\in \pathindices$.  We can then bound the complexity of the proposed  analysis by $\mathcal{O}(\mathit{fwra} + |\pathindices|\cdot \jumpmax\cdot\mathit{bwra})$, where 
 the number of sets used for numerical integration is in $\mathcal{O}(|\pathindices|\cdot \jumpmax)$.

\section{Feasibility study}\label{sec:casestudy}

\begin{figure}[t]%
\centering\begin{tikzpicture}[
n/.style={draw, text width = 1.3cm, minimum height = 1.7cm, align = center, font= \footnotesize, rounded corners, very thick, execute at begin node=\setlength{\baselineskip}{8pt}%
},
nsmall/.style={draw, text width = 1.4cm, minimum height = 1cm, align = center, font= \footnotesize, rounded corners, very thick, execute at begin node=\setlength{\baselineskip}{8pt}%
},
charging/.style={minimum height = 1.3cm
},
en/.style={draw=none, minimum height=0cm, font = \scriptsize, align = center},%
c/.style={draw, fill, black, circle, inner sep=0, outer sep=0, minimum size=1mm},
l/.style={anchor=west, inner sep=0, font=\footnotesize},
]

\node[nsmall,charging](chA) at (1.5,0) {
	$\text{charging}_A$	\\
	\vspace{0.125cm} 	
	$\dot{x}\in[3,6]$\\
	$x\in[0,3]$
	};

\node[nsmall,charging](chB) at (4.1,0) {
	$\text{charging}_B$\\
	\vspace{0.125cm} 	
	$\dot{x}\in[2,5]$\\
	$x\in[3,8]$	};

\node[nsmall,charging](chC) at (6.7,0) {
	$\text{charging}_C$	\\
	\vspace{0.125cm} 	
	$\dot{x}\in[1,2]$\\
	$x\in[8,10]$	};

\node[nsmall,charging](noch) at (9.3,0) {
	$\text{full}$	\\
	\vspace{0.125cm} 	
	$\dot{x}=0$\\
	$x=10$	};

\node[nsmall](dr) at ($(chA)-(0,2)$) {
	$\text{driving}$\\
	\vspace{0.125cm} 	
	$\dot{x}=-1$\\	};

\node[nsmall](arr) at (-0.9,-2) {
	$\text{arrival}$	};
	
	\node[nsmall](em) at  ($(chB)-(0,2)$)  {
	$\text{empty}$	
	};

\node[nsmall](re) at ($(chC)-(0,2)$) {
	$\text{detour}$\\
	\vspace{0.125cm} 	
	$\dot{x}=-1$
	};

\node[nsmall](new) at ($(noch)+(0,-2)$) { %
	$\text{charge}$\\
	};

\node[en, text width=1.2cm,execute at begin node=\setlength{\baselineskip}{8pt}, anchor=north] (init) at (-0.65,0.54) {$x\in${$\,[0,2]$}\\
$c=0$\\$d=0$\\
	${r}=0$ };
\draw[-latex, very thick] ($(chA.west)-(0.6,0)$) to  ($(chA.west)+(0,0)$);

\draw[-latex,  very thick] (chA) to node[en, above] {$x=3$} (chB);
\draw[-latex, very thick] (chB) to node[en, above] {$x=8$} (chC);
\draw[-latex, very thick] (chC) to node[en, above] {$x=10$} (noch);

\draw[-latex,  very thick] (chA) to node[en, left] {$c$} (dr);
\draw[-latex,  very thick] (chB.south) -- ($(chB)-(0,1)$) to %
($(chA)-(0,1)$)  -- (dr.north);
\draw[-latex,  very thick] (chC.south) -- ($(chC)-(0,1.1)$) to %
($(chA)-(0,1.1)$) -- (dr.north);
\draw[-latex,  very thick] (noch.south) -- ($(noch)-(0,1.2)$) to %
($(chA)-(0,1.2)$) -- (dr.north);

\draw[-latex,  very thick] (dr) to node[en, above] {$d$} (arr);
\draw[-latex,  very thick] ($(dr.south)+(0,0)$)  --  ($(dr.south)+(0,-0.3)$) to  node[en, above, near start, ] {$r$}($(re.south)+(0,-0.3)$) -- (re);
\draw[-latex,  very thick,] (dr) to node[en, above] {$x=0$} (em);

\draw[-latex,  very thick,] (re) to node[en, above] {$x=0$} (em);

\draw[-latex, very thick] (re)  to node[en, above] {   $d $}  ($(new.west)+(0,0)$);%

\draw[-latex,  very thick] ($(new.east)$) %
-- ($(new.east)+(0.2,0)$) %
-- ($(noch.north east)+(0.2,0.7)$) 
-- ($(chA.north)+(0,0.7)$) 
to node[en, left] {$x\in[0,3]$}
 (chA);

\draw[-latex, very thick] ($(new.east)$) %
-- ($(new.east)+(0.2,0)$) %
-- ($(noch.north east)+(0.2,0.7)$) 
-- ($(chB.north)+(0,0.69)$) 
to node[en, left] {$x\in[3,8]$} (chB);

\draw[-latex, very thick]  ($(new.east)$) %
-- ($(new.east)+(0.2,0)$) %
-- ($(noch.north east)+(0.2,0.7)$) 
-- ($(chC.north)+(0,0.685)$) 
to node[en, left] {$x\in[8,10]$} (chC);

\draw[-latex, very thick] ($(new.east)$) %
-- ($(new.east)+(0.2,0)$) %
-- ($(noch.north east)+(0.2,0.7)$) 
-- ($(chC.north)+(2.6,0.68)$) 
to node[en, left] {$x=10$} (noch);

\end{tikzpicture}
\caption{Car model with detours. Random clock $c$ is active ($\dot{c}=1$) in the charging locations, $d$ is active in locations \emph{driving} and \emph{detour} and $r$ is active in location \emph{driving}. The state of charge $x$ is restricted to $[0,10]$ in all locations unless stated otherwise. No time is spent in location \emph{charge} due to invariants not shown. %
}\label{fig:casestudy_loop}
\end{figure}
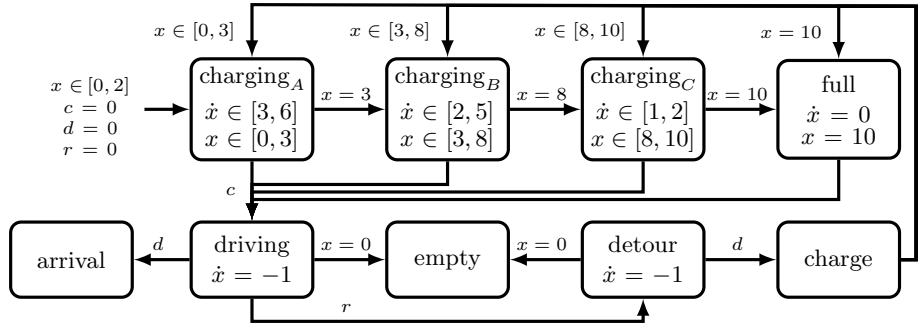

Figure~\ref{fig:casestudy_loop} illustrates our model of an electric car with different charging modes, which choose the charging rate nondeterministically from different intervals.
Charging stops whenever the battery is full. The driving distances are sampled from random variables. See~\cite{delicaris2023maximizing} %
for the  automaton with all dynamics.

Locations \emph{charging$_A$}, \emph{charging$_B$}, \emph{charging$_C$} and \emph{full} model  decreasing charging rates, depending on the state of charge of the battery. 
The charging time is modeled by random clock $c$ which follows a folded normal distribution, i.e. $\mathcal{N}(2,2)$. 
In location \emph{driving}, the battery decreases, where   random clock  $d$ models the time spent driving. 
The expiration of $d$ leads to location \emph{arrival}, while draining the battery leads to location \emph{empty}.
The driving time  follows a folded normal distribution ($\mathcal{N}(4,1)$).

The model is scalable, as it includes the possibility of taking $0$ or more detours. 
In location \emph{driving}, the random delay until the next detour ($r$)  competes with the end of the drive ($d$). 
Random clock $r$ follows an exponential distribution with $\lambda=2$.
In the \emph{detour} location, $d$ is still active and the battery still decreases.
The expiration of $d$ corresponds to the end of the detour and the transition to location \emph{charge} is taken, which  marks the start of the next charging cycle. %
Depending on the current charge of the battery, the transition to the matching charging location is chosen immediately. 
By explicitly restricting the number of detours, we ensure that the model is Zeno-free.
Performing a worst case analysis, we compute the maximum probability  of reaching an \emph{empty} battery. %

\paragraph{Implementation and reproducibility.}
We rely on  a prototypical implementation of the tool \realyst%
\footnote{Tool and models: \url{https://zivgitlab.uni-muenster.de/ag-sks/tools/realyst/}.} 
and  use the library \hypro \cite{schupp2017HyProLibraryState} to compute flowpipe segments; particularly \hypro is used to compute the forward flowpipe, the reach tree, and the backward time closure and  one-step relation as used in our backward refinement. 
Additionally, we use \gsl~\cite{gsl} to perform multi-dimensional integration. %
All experiments were run on a machine equipped with an Intel\smash{\textsuperscript{\textregistered} Core\textsuperscript{\texttrademark}} i7 with 6$\times$\SI{3.30}{\giga\hertz} and \SI{32}{\giga\byte} RAM.

\paragraph{Results.}

\begin{table*}[bt]
    \centering
    \scriptsize
    \caption{
      Max reachability probabilities  for  $\Goal=\{(\text{empty}, \val) \mid \val_x =0\}$, $\tmax = (\#\text{detours}+1)\cdot 21$ and $\#\text{detours} \in \{0,1,2\}$.
   Computation times $\comptime$ provided for \realyst (\realystshort) with error $\estat$ and \prohver (\prohvershort). %
    $K=(|\VarRandom|, |\reachtree|, |I_S|)$. %
      }
    \label{tab:results}
    \newcolumntype{Y}{>{\centering\arraybackslash}X}
    \renewcommand{\arraystretch}{1}
    \begin{tabularx}{\linewidth}{lp{0.3cm}p{0.55cm}YYYYYYY}
        \toprule
        \multicolumn{2}{c}{\ }   & \   &    \multicolumn{3}{c}{$0$ detours} &  \multicolumn{3}{c}{$1$ detour}  & $2$ detours  \\
      \cmidrule(lr){4-6}\cmidrule(lr){7-9}\cmidrule(lr){10-10}
            \multicolumn{2}{c}{\ }     & var.    & \texttt{A}& \texttt{AB}& \texttt{ABC}& \texttt{A} &\texttt{AB}& \texttt{ABC}&  \texttt{A}   \\
      \multicolumn{2}{c}{\ }   & $K$ & $(2,8,2)$ &  $(2,12,3)$ & $(2,16,4)$  & $(5,38,10)$ & $(5,88,19)$  & $(5,167,39)$    &  $(8,128,34)$   \\
          \midrule
          \multirow{6}{*}{\rotatebox{90}{rectangular}} 
            & \multirow{3}{*}{\shortstack[l]{\realystshort}} %
            & $p_\textsl{max}$  
            & \probab{0.0503536} & \probab{0.0648239} & \probab{0.0666976} & \probab{0.407864} & \probab{0.42621} & \probab{0.429555} & \probab{0.402105} \\ 
            & & $\estat$ 
            & \errortable{8.76307e-05} & \errortable{0.000119219} & \errortable{0.000128606} & \errortable{0.00171464} & \errortable{0.00167712} & \errortable{0.000613131} & \errortable{0.00332672} \\
            & & {$\comptime$}  
            & \rt{4.49372} & \rt{10.745} & \rt{15.3086} & \rt{484.5} & \rt{1166.49} & \rt{1800.63} & \rt{35570.3} \\
            \cmidrule{2-10} 
            & \multirow{2}{*}{\shortstack[l]{\prohvershort}} %
            & $p_\textsl{max}$  
            & \probab{0.3055917831257279} & \probab{0.32760017561658367} & \probab{0.32760017561658367} & \probab{0.5799221576213734} & \probab{0.6145757767099783} & \probab{0.590174399844356} & \probab{0.7949767298910101} \\
            & & {$\comptime$}  
            & \rt{613.475} & \rt{1202.655} & \rt{1735.179} & \rt{7890.288} & \rt{15247.254} & \rt{28533.796} & \rt{1881.607} \\
          \midrule
          \multirow{6}{*}{\rotatebox{90}{singular}} 
            & \multirow{3}{*}{\shortstack[l]{\realystshort}} %
            & $p_\textsl{max}$  
            & \probab{0.0266728} & \probab{0.0265343} & \probab{0.0265998} & \probab{0.406624} & \probab{0.424759} & \probab{0.429528} & \probab{0.408268} \\
            & & $\estat$
            & \errortable{3.40482e-05} & \errortable{9.47598e-05} & \errortable{0.000112267} & \errortable{0.000711971} & \errortable{0.000670252} & \errortable{0.00167624} & \errortable{0.0037186} \\
            & & {$\comptime$}  
            & \rt{2.46578} & \rt{11.6589} & \rt{19.3762} & \rt{245.213} & \rt{518.298} & \rt{1720.16} & \rt{12187.8} \\
            \cmidrule{2-10} 
            & \multirow{2}{*}{\shortstack[l]{\prohvershort}} %
            & $p_\textsl{max}$  
            & \probab{0.3055917831257279} & \probab{0.32760017561658367} & \probab{0.32760017561658367} & \probab{0.44295903375961115} & \probab{0.4679160428204017} & \probab{0.5239576415909519} & \probab{0.7580344451394565} \\
            & & {$\comptime$}  
            & \rt{348.091} & \rt{500.185} & \rt{546.559} & \rt{6984.118} & \rt{10249.278} & \rt{14273.85} & \rt{1461.853} \\
          \bottomrule
          \hspace{0.1cm}
    \end{tabularx}
\end{table*}

We consider three different model variants, namely \texttt{ABC}, \texttt{AB} and \texttt{A}, that include exactly those charging locations indicated in the model name. 
If locations \emph{charging$_B$} and/or \emph{charging$_C$} are removed, the invariant in the last charging location is extended to $x\leq 10$ and the in- and outgoing transitions of that location are adapted accordingly.   
To further reduce model complexity, a singular version is created, where the rate of the continuous variables equals the lower bound of the corresponding rectangular interval. 
Continuous nondeterminism is maintained via the initial set, hence results cannot be compared with \cite{Pilch2021Optimizing}.%

Table~\ref{tab:results} contains maximum reachability probabilities and the corresponding computation time for the full model presented in Figure~\ref{fig:casestudy_loop} (indicated by \texttt{ABC}) and for reduced model versions (\texttt{AB} and \texttt{A}).
We scale $\tmax = (\#\text{detours}+1)\cdot 21$, for $\#\text{detours} \in \{0,1,2\}$ %
and use \samples{1e5}, \samples{2e6}, \samples{1e7} integration samples for $0,1,2$ detours.
We validate results for different parameter settings computed by our prototype \realyst with results from \prohver~\cite{hahn2013CompositionalModellingAnalysis},
which computes a safe overapproximation of the reachability probabilities via discretization.

Table~\ref{tab:results} indicates for every considered number of detours the resulting size of the model as tuple $\smash{K=(|\VarRandom|, |\reachtree|, |\Goalreached|)}$, where $|\VarRandom|$ is the number of random variables, $|\reachtree|$ the number of nodes in the reach tree and $|I_S|=|\Goalreached|$ the number of traces leading to the goal set. 
The dimension of the polytopes constructed by forward and backward analysis is $|\VarRandom|+3$, and the dimension of integration equals $|\VarRandom|$.

 \realyst  computes maximum reachability probabilities for all model variants with $0$ and $1$ detour in at most $30$ minutes.
For $2$ detours, the complexity of the model increases considerably.
Computations in the singular model \texttt{A} take up to $3.5$ hours, and in  the rectangular variant  for $2$ detours just below $10$ hours.
The number of dimensions in variants \texttt{AB} and \texttt{ABC} with $2$ detours becomes too large, such that flowpipe construction does not terminate or takes very long.  
\realyst is able to complete the singular variant of \texttt{AB}  in slightly less than \SI{83}{\hour} and results in probability $\probab{0.431565}$ with  $\estat=\error{0.00706381}$ statistical error.

The probability to reach an empty battery increases considerably with additional detours, as they
 introduce uncertainty to the state of charge of the battery. The scheduler can exploit this to maximize the probability of an empty battery. 

Maximizing the reachability probability for an undesired goal yields a \emph{worst case} probability.
Reaching an empty battery is undesirable, hence, the computed probability to reach the goal provides an upper bound when everything goes wrong.
The results indicate that modeling the charging process in more detail has a relatively low impact on the computed probability. 
This is expected, as the influence of charging on the state of charge of the battery (rates between $1$ and $6$) is in any case higher than the influence of driving on the state of charge of the battery (rate $-1$).
Rectangular behavior gives a scheduler plenty opportunities to impact model evolution, which may increase the reachability probability.

As of $1$ detour, the results for rectangular and singular models are  close.
This is due to  the singular rates being equal to the lower bounds of the rectangular rate intervals. 
The scheduler aims to reduce the state of charge of the battery, hence, in most cases it will choose  the lowest possible rate.

\prohver computes safe overapproximations of maximum reachability probabilities.
However, its precision highly depends on the chosen number of discretization intervals.
A recent release of \prohver automates the interval generation as well as a refinement thereof, w.r.t. to given parameters.
For $0$ detours, \prohver  computes  a substantial overapproximation of the reachability probability obtained by \realyst. 
Computation times of \prohver are  between $28$ and $140$ times larger.   
For $1$ detour, \prohver  takes up to \SI{8}{\hour} and results in a better approximation of the reachability probabilities. 
For $2$ detours, \prohver is not able to perform a refinement on its discretization, yielding quick computation times with a substantial overapproximation. 
Running \prohver with alternative parameters which enforce more discretization intervals does not terminate in less than \SI{15}{\hour}.
We refer to \ref{appendix:prohver} for details on the parameter setting of \prohver; and to \ref{appendix:comptimes} for details on the computation times of \realyst.

\realyst indicates an error between \scientific{e-5} up to \scientific{e-3}, which  due to Lemma~\ref{lemma:union}, solely stems from integration.
For the choice of $\tint= 100$ and the distributions $\mathcal{N}(4,1)$,  $\mathcal{N}(2,2)$ and $Exp(2)$, the computed error $\einfty=0$, using IEEE 754 double precision. %
Hence, the probabilities computed by \realyst plus the indicated error $\estat$ agree with the overapproximations provided by \prohver.

\section{Conclusions and Future Work}\label{sec:conclusion}
 \reviewstatus
 We propose rectangular automata with random clocks as a new modeling formalism that combines discrete and continuous behavior with random delays.
Nondeterminism is usually resolved probabilistically in stochastic hybrid systems. 
In contrast, this paper presents the first approach to compute maximum reachability probabilities for rectangular automata with random clocks, fully resolving all kinds of discrete and continuous nondeterminism prophetically. 
The computation requires a combination of forward flowpipe construction with a backward refinement to partition the potentially infinite set of schedulers. 
The resulting error solely stems from the multidimensional integration. 
The results of the feasibility study show that \realyst performs very well for up to five random variables.
Reachability probabilities are highly accurate and obtained fast in comparison to \prohver. %
Future work aims to improve scalability via other state set representations, and  compute  prophetic minimum reachability probabilities. 
We will provide an equivalent notion of RAR where restrictions on random delays are placed implicitly via the semantics to ease modeling; a transformation between both will maintain analyzability via \realyst. %

\bibliographystyle{splncs04}
\bibliography{references}

\begin{appendix}
\renewcommand{\thesection}{Appendix \Alph{section}}
\label{appendix}

\section{Proofs}\label{appendix:proofs}
\setcounter{lemma}{0} \setcounter{equation}{1}
\begin{lemma}\label{appendix:lemma:rates}
Assume $\hastate_0\xrightarrow{t_1,\textit{rate}_1}\hastate_1\xrightarrow{t_2,\textit{rate}_2}\hastate_2$ for $\hastate_0,\hastate_1,\hastate_2\in\States$ with location $\ell$, $t_1,t_2\in\Realsposzero$  and $\textit{rate}_1,\textit{rate}_2 \in \Act(\ell)$. Then there is  $\textit{rate}\in \Act(\ell)$ s.t. $\hastate_0\xrightarrow{t_1 + t_2,\textit{rate}}\hastate_2$.
\end{lemma}

\begin{proof}[Correctness of Lemma~\ref{lemma:rates}]
Let  $\hastate_i=(\ell,\val_i,\valRandom_i)\in\Sigma$ for $i=0,1,2$, and let $t_1,t_2\in\Realsposzero$
and $\textit{rate}_1,\textit{rate}_2 \in \Act(\ell)$
with $\hastate_0\xrightarrow{t_1,\textit{rate}_1}\hastate_1\xrightarrow{t_2,\textit{rate}_2}\hastate_2$. From the semantics of time delay we know that
\begin{itemize}
\item there exists $\textit{rate}_1\in \Act(\ell)$ such that $\val_1=\val_0+t_1\cdot\textit{rate}_1$, $\valRandom_1=\valRandom_0+t_1\cdot\ActRandom(\ell)$, and
\item there exists $\textit{rate}_2\in \Act(\ell)$ such that $\val_2=\val_1+t_2\cdot\textit{rate}_2$, $\valRandom_2=\valRandom_1+t_2\cdot\ActRandom(\ell)$, and for all $\clocki \in \VarRandom$ there exists some $v \in \support(\Distr(\clocki))$ such that ${\valRandom_2}_\clocki \leq v$.
\end{itemize}
The case $t_1+t_2=0$ is straightforward. For $t_1+t_2>0$ we observe
\begin{eqnarray*}
\val_2 & = & \val_0+t_1\cdot\textit{rate}_1+t_2\cdot\textit{rate}_2\\
&=& \val_0+(t_1+t_2)\cdot\frac{t_1\cdot\textit{rate}_1+t_2\cdot\textit{rate}_2}{t_1+t_2}\\
&=& \val_0+(t_1+t_2)\cdot\big[
\frac{t_1}{t_1+t_2}\cdot\textit{rate}_1+
\frac{t_2}{t_1+t_2}\cdot\textit{rate}_2\big]\ .
\end{eqnarray*}
Since $t_1$ and $t_2$ are non-negative, also the factors $\frac{t_1}{t_1+t_2}$ and $\frac{t_2}{t_1+t_2}$ are non-negative and sum up to 1, i.e. the linear combination $\textit{rate}=\frac{t_1}{t_1+t_2}\cdot\textit{rate}_1+\frac{t_2}{t_1+t_2}\cdot\textit{rate}_2$ of the rates $\textit{rate}_1$ and $\textit{rate}_2$ from the convex set $\Act(\ell)$ is also included in $\Act(\ell)$. With
\begin{eqnarray*}
\valRandom_2&=&\valRandom_1+t_2\cdot\ActRandom(\ell)\\
&=&\valRandom_0+t_1\cdot\ActRandom(\ell) + t_2\cdot\ActRandom(\ell)\\
&=&\valRandom_0+(t_1+t_2)\cdot\ActRandom(\ell)
\end{eqnarray*}
we have shown that all antecedents of the time step $\hastate_0\xrightarrow{t_1+t_2,\textit{rate}}\hastate_2$ are satisfied.
\end{proof}

In the following we prove the correctness of the derived sample domain for a maximizing prophetic scheduler, i.e.
the equality of $\bigcup_{\scheduler\in\SchedulersProphMax}\valSet^\scheduler=\Polymax$: 
\begin{lemma}\label{appendix:lemma:union}
Given the set of valuations $\valSet^\scheduler$ over the space of the random variables that allow scheduler $\scheduler$ to reach the goal set $\Goal$,
and an equivalence relation over $\SchedulersProphMax$ defining equivalence classes $\SchedulersRepr$, 
$\bigcup_{\scheduler\in\SchedulersProphMax}\valSet^\scheduler$ can be computed as follows:
\begin{align} \label{eq:integrationdomain}
    \bigcup_{\scheduler\in\SchedulersProphMax}\valSet^\scheduler 
    &= \bigcup_{i \in I_S}\bigcup_{\scheduler \in \SchedulersRepr}\valSet^\scheduler
    = \bigcup_{i \in I_S} \valSet ^ \SchedulersRepr
    = \bigcup_{i \in I_S} \Polymaxi
    =\Polymax,
\end{align}
 where
 $\valSet^\SchedulersRepr$ is the %
 union of valuations induced by  schedulers $\scheduler\in \SchedulersRepr$.
\end{lemma}

Recall that, in the extraction of the sample domain (c.f. Section~\ref{subsec:collection}), we iteratively traverse each trace $i$ leading to a goal state set $\States^i$, collecting constraints on the samples. 
Here, we distinguish in each segment if a random delay is i) already expired, ii) about to expire by taking the next step in the reach tree or iii) not expired.
Hence, on each trace, there is one segment $\RefinedSegmentik{i}{k}$ for every random delay that represents the valuation set where this random delay is about to expire.
We map the index $k$ to the corresponding random clock $\clocki_n \in \VarRandom$ with $\map: I_i \rightarrow \VarRandom$, where $I_{i}$ collects the indices $k$ of all such segments $\RefinedSegmentik{i}{k}$.
Note that, $|I_{i}|\leq|\VarRandom|$, as for all random clocks that have not yet expired in $\Goali=(l,\Goalvali)$, no such segment exists.

We then use $P^i_k$ with $i\in I_S$ and $ k \in I_i$ to denote the state of $\Polymaxi$ after deriving constraints from $\RefinedSegmentik{i}{k}$.
Consequently, for each random delay $\clocki_n$, the polytope $P^i_k$ with $\map(k)=\clocki_n$ contains the upper bounds for the sample $\sample_{\clocki_n}$.
We can then describe the computation of $\Polymaxi$ as an intersection of all $P^i_k$:
$ \Polymaxi = \left( \bigcap_{k \in I_i} P^i_k \right) \cap P^i $; where we use $P^i$ without index $k$ to denote the constraints that can be derived from $\valSet^i$ itself. %
We use $\Project{\{r\}}{P}$ to denote the projection of a polytope $P$ onto dimension $r$.
Using these notations, we can prove Lemma~\ref{lemma:union}.

\begin{proof}[Correctness of Lemma~\ref{lemma:union}]
In the presence of continuous nondeterminism, the set of the maximizing schedulers $\SchedulersProphMax$ is uncountably infinite.
We use that a partition of schedulers leading to the same state set $\Goali \subset \Goal$ via one trace and reach tree (which we assigned the trace index $i$ to) can be considered collectively by the equivalence class $\SchedulersRepr$.
Hence, taking the union over all $\SchedulersRepr$ and all schedulers within every equivalence class in Equation~\ref{eq:integrationdomain},  fully covers the set of maximizing schedulers $\SchedulersProphMax$. 
Consequently, the first equality in Equation~\ref{eq:integrationdomain} follows from the properties of an equivalence relation and the second equality follows simply from the definition of $\valSet^{\SchedulersRepr}$.
The fourth equality follows from the definition of $\Polymax$,
which is the result of the proposed backward analysis approach, outlined in Section~\ref{subsec:refinement} and \ref{subsec:collection}.

For the third equality, we argue over the trace index $i$.
Each set $\SchedulersRepr$ corresponds to a trace $i$ induced by the schedulers from $\SchedulersRepr$.
For each trace $i$, $\Polymaxi$ defines constraints on  all samples, such that schedulers with the ability to make decisions leading to $\Goali$ exist.
The third equality hence stems from the equality $\valSet ^\SchedulersRepr=\Polymaxi$ for all $i\in \pathindices$, which we show in the following:
  \begingroup
\allowdisplaybreaks
         \begin{align*}%
         &\sample \in \Polymaxi = \left( \bigcap_{k \in I_i} P^i_k \right) \cap P^i 
         \label{eq:proof1} \tag{1}\\
         \Leftrightarrow & \ 
         \sample \in P^i
         \land
         \forall k \in I_i: \sample \in P^i_k 
         \label{eq:proof2} \tag{2}\\
         \Leftrightarrow & \ \forall r_n \in \VarRandom :
         \big(
         \sample_{r_n} \in \Project{\{r_n\}}{P^i}
         \land
         \forall k \in I_i: \sample_{r_n} \in \Project{\{r_n\}}{P^i_k}
         \big)
         \label{eq:proof3} \tag{3}\\
         \Leftrightarrow & \ \forall r_n \in \VarRandom :
         \forall k \in I_i: \big(
         \sample_{r_n} \in \Project{\{r_n\}}{P^i} 
         \land 
         \sample_{r_n} \in \Project{\{r_n\}}{P^i_k}
         \big)
         \label{eq:proof4} \tag{4}\\
         \Leftrightarrow & \ \forall r_n \in \VarRandom :
         \forall k \in I_i: 
         \label{eq:proof5} \tag{5}\\
         &\  \big(
         \sample_{r_n} \in  \Project{\{r_n\}}{P^i} 
         \land 
         \sample_{r_n} \in  \Project{\{r_n\}}{P^i_k} \land \map(k)=\clocki_n 
         \big)\\
         &\ \lor \big(
         \sample_{r_n} \in  \Project{\{r_n\}}{P^i}
         \land 
         \sample_{r_n} \in  \Project{\{r_n\}}{P^i_k} \land \map(k)\not =\clocki_n 
         \big)
         \\
         \Leftrightarrow & \ \forall r_n \in \VarRandom :
         \label{eq:proof6} \tag{6}\\
         &\  \Big( \forall k \in  \{k \in I_i \mid \map(k)  =\clocki_n \}: 
          \big( 
         \sample_{r_n} \in  \Project{\{r_n\}}{P^i}
         \land 
         \sample_{r_n} \in  \Project{\{r_n\}}{P^i_k} 
         \big)\Big)\\
         &\ \land \Big( \forall k \in  \{k \in I_i \mid \map(k) \not =\clocki_n \}: \big( 
         \sample_{r_n} \in  \Project{\{r_n\}}{P^i}
         \land 
         \sample_{r_n} \in  \Project{\{r_n\}}{P^i_k} 
         \big)\Big)
         \\
         \Leftrightarrow & \ \forall r_n \in \VarRandom :  
         \label{eq:proof7} \tag{7}\\
         &\  \Big( \exists k \in I_i: \map(k)=\clocki_n  \Rightarrow \big(  \sample_{r_n} \in \Project{\{r_n\}}{P^i} 
         \land 
         \sample_{r_n} \in \Project{\{r_n\}}{P^i_k}
         \big)\Big)\\
         &\ \land \Big( \forall k \in  \{k \in I_i \mid \map(k) \not =\clocki_n \}: \big( 
         \sample_{r_n} \in \Project{\{r_n\}}{P^i}
         \land 
         \sample_{r_n} \in \Project{\{r_n\}}{P^i_k} 
         \big)\Big)
         \\
         \Leftrightarrow & \ \forall r_n \in \VarRandom: 
         \Big(\sample_{r_n} \in \Project{\{r_n\}}{P^i}
         \land 
         \label{eq:proof8} \tag{8}\\
         &\ \big(\exists k \in I_i: (\map(k)=\clocki_n) \Rightarrow\sample_{r_n} \in \Project{\{r_n\}}{P^i_k} \big)\Big)
         \\
        \Leftrightarrow &\ \forall r_n \in \VarRandom :  
         \Big(\sample_{r_n} \in \Project{\{r_n\}}{\Lift{\{\contrandomvar\}}{\Goalvali}}   \label{eq:proof9} \tag{9}\\
        &\  \land \big(\exists k \in I_i: (\map(k)=\clocki_n) \Rightarrow \sample_{r_n} \in \Project{\{r_n\}}{\RefinedSegmentik{i}{k}}\big)   \Big)  
        \\
         \Leftrightarrow &\ \exists \scheduler \in \SchedulersRepr : \exists \hahistory \text{ initial } : \hahistory = \hastate_0 \rightarrow \dots \rightarrow \hastate_n \land  \hastate_n \in \Goali\subset \Goal \land  \label{eq:proof10} \tag{10}\\
         & \ \forall r_n \in \VarRandom:  \exists b \in \{0, \dots, n\} : \ \\ 
         &\  \Big( b=n\lor \big(\exists k \in I_i: (\map(k)=\clocki_n) \Rightarrow b=n-2\cdot k \big) \Big)  \\
         &\  \land \forall \hastate_a= (\ell_a, \val_a, \valRandom_a, \sample_a) \in \{\hastate_0, \dots, \hastate_b\}:\sample_a =\sample_{r_n} 
        \\     
         \Leftrightarrow &\ \sample \in \valSet^\SchedulersRepr
         \label{eq:proof11} \tag{11}
    \end{align*}
    \endgroup
The first two equivalences (\ref{eq:proof1} $\Leftrightarrow$ \ref{eq:proof2} $\Leftrightarrow$ \ref{eq:proof3}) follow from the definitions provided in Section~\ref{subsec:collection}. 
The next equivalences (\ref{eq:proof3} $\Leftrightarrow$\ref{eq:proof4} $\Leftrightarrow$ \ref{eq:proof5}$\Leftrightarrow$ \ref{eq:proof6}) distinguish over the indices $k$ which map to a certain variable $r_n$ and those which don't.
Specifically, since for each $r_n$, there exists at most one $k$ with $\map(k)=r_n$, the equivalence \ref{eq:proof6}$\Leftrightarrow$ \ref{eq:proof7} holds.

While implications \ref{eq:proof7} $\Rightarrow$ \ref{eq:proof8} $\Rightarrow$ \ref{eq:proof9} are straightforward, the implications \ref{eq:proof9} $\Rightarrow$ \ref{eq:proof8} $\Rightarrow$ \ref{eq:proof7} hold, since the additional segments $P^i_k$ and $P^i$ do not pose stronger restrictions on the valuations due to lifting and to random clocks being stopwatches, i.e. the valuations are either increasing or not changing with progression in the automaton. 
Note that, for all variables $r_n$ with $\map(k)=r_n$, all $P^i_{k'}$ with $k'\not = k$ only contain lower bounds, where  (i) for $P^i_{k'}$  with $k'< k$ they are less restrictive than the lower bounds in $P^i_k$, since $r_n$ is not yet expired, and (ii) for $P^i_{k'}$  with $k'> k$ (and analogously $P^i$) the lower bound is equal to the lower bound obtained from $P^i_k$, due to expiration.
For $r_n$ that do not expire on trace $i$, (i) holds for every $P^i_{k}$, since $ \forall k \in I_i: \map(k)\not =r_n$.

Since segments $\RefinedSegmentik{i}{k}$ and $\Goalvali$ are part of the forward flowpipe (and backward refinement), it is possible to find a scheduler following that trace with valuations $\sample \in \Polymaxi$ (\ref{eq:proof9} $\Leftrightarrow$ \ref{eq:proof10}). 
Hence, an initial path exists, where the sample values until the first (and only) expiration match $\sample$. 
This corresponds to knowledge on future expiration times of random variables. Scheduling decisions are then taken, based on this knowledge, i.e. prophetically. 
The last equivalence (\ref{eq:proof10} $\Leftrightarrow$ \ref{eq:proof11}) follows from the definition of $ \valSet^\SchedulersRepr$.

Termination of the proposed method is given due to the finiteness of the reach tree. The backward refinement is computed over all segments $\Goalvali$, each corresponding to one segment in the reach tree. The computation then follows the trace to the root of the reach tree until the root is reached.
Combining the relevant segments on each trace is then also bounded due to finiteness of the traces.
\end{proof}

\section{Running example computations}\label{appendix:runningexample}

The refined segment $\RefinedSegmentik{0}{2}$ and the next intermediate goal segment $\JumpSegmentik{0}{3}$  (c.f. Figure~\ref{fig:backward_ref_steps0}) can be computed from $\JumpSegmentik{0}{2}$, resulting in the following segments:

\begin{align*}
    \JumpSegmentik{0}{2} %
    &= \{ \val \in \mathbb{R}^3 \mid \val_x = 4 \land \val_y \leq \val_x \land \val_y \geq \val_x -1 \land \val_r=3\}\\
    \hat{\valSet}^0_2 &=  \btc{\ell_0}( \JumpSegmentik{0}{2} ) \cap \valSet^0_2 = %
    \{ \val \in \mathbb{R}^3 \mid \val_x \leq 4
    \land \val_y \geq \nicefrac{5}{3} + \nicefrac{1}{3}\cdot \val_x  
    \\& 
\land \val_y \leq \nicefrac{8}{3} + \nicefrac{1}{3}\cdot \val_x  
    \land \val_r = \val_x-1\}\\
    \JumpSegmentik{0}{3} &= \hat{\valSet}^0_2 \cap \Init(\ell_0)%
    = \{ \val \in \mathbb{R}^3 \mid \val_x \leq 1 \land \val_y \in [2,3] \land \val_r = 0 \}
\end{align*}

The refined segment $\RefinedSegmentik{1}{1}$ and the next intermediate goal segment $\JumpSegmentik{1}{2}$ (c.f. Figure~\ref{fig:backward_ref_steps1}) are computed from $\JumpSegmentik{1}{1}$, resulting in the following segments:
\begin{align*}
    \JumpSegmentik{1}{1}
    &= \{ \val \in \mathbb{R}^3 \mid \val_x \in [2,4] \land \val_y \in [10,11] \land \val_r= \val_x -1\}\\
    \hat{\valSet}^1_1 &=  \btc{\ell_3}( \JumpSegmentik{1}{1} ) \cap \valSet^1_1 %
    =\{ \val \in \mathbb{R}^3 \mid  \val_x \in [2,4] \land \val_y \in [1,11] \land \val_r= \val_x -1\}\\
    \JumpSegmentik{1}{2} &= \bosr{e}(\hat{\valSet}^1_2) \cap \valSet^1_2
    =\{ \val \in \mathbb{R}^3 \mid  \val_x \in [2,4] \land \val_y \geq 1 
    \\&
    \land \val_y \leq  \nicefrac{8}{3} + \nicefrac{1}{3}\cdot \val_x  \land \val_r= \val_x -1\}
\end{align*}

\section{Feasibility study}\label{appendix:casestudy}
\subsection{Car model}\label{appendix:model}
 
 \begin{figure*}[h]%
     \centering
     \begin{tikzpicture}[
n/.style={draw, text width = 1.45cm, minimum height = 3.3cm, align = center, font= \footnotesize, rounded corners, very thick, execute at begin node=\setlength{\baselineskip}{8pt}%
},
nsmall/.style={draw, text width = 1.5cm, minimum height = 1cm, align = center, font= \footnotesize, rounded corners, very thick, execute at begin node=\setlength{\baselineskip}{8pt}%
},
en/.style={draw=none, minimum height=0cm, font = \scriptsize, align = center},%
c/.style={draw, fill, black, circle, inner sep=0, outer sep=0, minimum size=1mm},
l/.style={anchor=west, inner sep=0, font=\footnotesize},
]

\node[n](chA) at (1.5,0) {
	$\text{charging}_A$	\\
	\vspace{0.2cm} 	
	$\dot{t}=1$\\
	$\dot{x}\in[3,6]$\\
	$\dot{q}=0$\\
	$\dot{c}=1$ \\ 
	$\dot{d}=0$ \\ 
	$\dot{r}=0$ \\ 
	\vspace{0.2cm} 
	$x\in[0,3]$
	};

\node[n](chB) at (4.2,0) {
	$\text{charging}_B$\\
	\vspace{0.2cm} 	
	$\dot{t}=1$\\
	$\dot{x}\in[2,5]$\\
	$\dot{q}=0$\\	
	$\dot{c}=1$ \\ 
	$\dot{d}=0$ \\ 
	$\dot{r}=0$ \\ 
	\vspace{0.2cm} 
	$x\in[3,8]$	};

\node[n](chC) at (6.9,0) {
	$\text{charging}_C$	\\
	\vspace{0.2cm} 	
	$\dot{t}=1$\\
	$\dot{x}\in[1,2]$\\	
	$\dot{q}=0$\\
	$\dot{c}=1$ \\ 
	$\dot{d}=0$ \\ 
	$\dot{r}=0$ \\ 
	\vspace{0.2cm} 
	$x\in[8,10]$	};

\node[n](noch) at (9.6,0) {
	$\text{full}$	\\
	\vspace{0.2cm} 	
	$\dot{t}=1$\\
	$\dot{x}=0$\\	
	$\dot{q}=0$\\
	$\dot{c}=1$ \\ 
	$\dot{d}=0$ \\ 
	$\dot{r}=0$ \\ 
	\vspace{0.2cm} 
	$x=10$	};

\node[n](dr) at ($(chA)-(0,4.2)$) {
	$\text{driving}$\\
	\vspace{0.2cm} 	
	$\dot{t}=1$\\
	$\dot{x}=-1$\\
	$\dot{q}=0$\\ 	
	$\dot{c}=0$ \\ 
	$\dot{d}=1$ \\ 
	$\dot{r}=1$ \\ 
	\vspace{0.2cm} 
	$x\in[0,10]$};

\node[n](arr) at ($(chC)-(0,4.2)$) {
	$\text{arrival}$\\
	\vspace{0.2cm} 	
	$\dot{t}=1$\\
	$\dot{x}=0$\\ 
	$\dot{q}=0$\\	
	$\dot{c}=0$ \\ 
	$\dot{d}=0$ \\ 
	$\dot{r}=0$ \\
	\phantom{$x\in[0,10]$}
	};
	
	\node[n,double](em) at  ($(noch)-(0,4.2)$)  {
	$\text{empty}$	\\
	\vspace{0.2cm} 	
	$\dot{t}=1$\\
	$\dot{x}=0$\\
	$\dot{q}=0$\\ 	
	$\dot{c}=0$ \\ 
	$\dot{d}=0$ \\ 
	$\dot{r}=0$ \\
	\phantom{$x\in[0,10]$}
	};

\node[n](re) at ($(dr)-(0,4.2)$) {
	$\text{detour}$\\
	\vspace{0.2cm} 	
	$\dot{t}=1$\\
	$\dot{x}=-1$ \\
	$\dot{q}=0$\\	
	$\dot{c}=0$ \\ 
	$\dot{d}=1$ \\ 
	$\dot{r}=0$ \\ 
	\vspace{0.2cm} 
	$x\in[0,10]$
	};

\node[n](new) at ($(arr)-(0,4.2)$) { %
	$\text{charge}$\\
 	\vspace{0.2cm} 	
 	$\dot{t}=1$\\
 	$\dot{x}=0$ \\	
 	$\dot{q}=1$ \\	
 	$\dot{c}=0$ \\ 
 	$\dot{d}=0$ \\ 
 	$\dot{r}=0$ \\ 
 	\vspace{0.2cm} 
	$x\in[0,10]$
	$q\leq0$		
	};

\node[en, text width=1.2cm,execute at begin node=\setlength{\baselineskip}{8pt}, anchor=north] (init) at (-0.65,0.54) {
$t=0$\\
$x\in${$\,[0,2]$}\\
$q=0$\\
$c=0$\\$d=0$\\
	${r}=0$ };
\draw[-latex, very thick] ($(chA.west)-(0.6,0)$) to  ($(chA.west)+(0,0)$);

\draw[-latex,  very thick] (chA) to node[en, above] {$x=3$} (chB);
\draw[-latex, very thick] (chB) to node[en, above] {$x=8$} (chC);
\draw[-latex, very thick] (chC) to node[en, above] {$x=10$} (noch);

\draw[-latex,  very thick] (chA) to node[en, left] {$c$} (dr);
\draw[-latex,  very thick] (chB.south) -- ($(chB.south)-(0,0.3)$) to %
($(chA.south)-(0,0.3)$)  -- (dr.north);
\draw[-latex,  very thick] (chC.south) -- ($(chC.south)-(0,0.4)$) to %
($(chA.south)-(0,0.4)$) -- (dr.north);
\draw[-latex,  very thick] (noch.south) -- ($(noch.south)-(0,0.5)$) to %
($(chA.south)-(0,0.5)$) -- (dr.north);

\draw[-latex,  very thick] (dr) to node[en, above] {$d$} (arr);
\draw[-latex,  very thick,] ($(dr.south)-(0.1,0)$) to node[en, left] {$r$} ($(re.north)-(0.1,0)$);
\draw[-latex,  very thick] ($(dr.south)+(0.1,0)$)  -- ($(dr.south)+(0.1,-0.4)$) to node[en, above, near start] {$x=0$} ($(em.south)-(0,0.4)$) -- (em);

\draw[-latex,  very thick] ($(re.north)+(0.1,0)$)  -- ($(dr.south)+(0.1,-0.5)$) to node[en, below, near start] {$x=0$} ($(em.south)-(0,0.5)$) -- (em);

\draw[-latex, very thick] (re)  to node[en, below] { $d$}  ($(new.west)+(0,0)$);%

\draw[-latex,  very thick] ($(new.east)$) %
-- ($(new.east)+(2.8,0)$) %
-- ($(chC.north east)+(2.8,0.7)$) 
-- ($(chA.north)+(0,0.7)$) 
to node[en, left] {$x\in[0,3]$}
 (chA);

\draw[-latex, very thick] ($(new.east)$) %
-- ($(new.east)+(2.8,0)$) %
-- ($(chC.north east)+(2.8,0.7)$) 
-- ($(chB.north)+(0,0.7)$) 
to node[en, left] {$x\in[3,8]$} (chB);

\draw[-latex, very thick]  ($(new.east)$) %
-- ($(new.east)+(2.8,0)$) %
-- ($(chC.north east)+(2.8,0.7)$) 
-- ($(chC.north)+(0,0.7)$) 
to node[en, left] {$x\in[8,10]$} (chC);

\draw[-latex, very thick] ($(new.east)$) %
-- ($(new.east)+(2.8,0)$) %
-- ($(chC.north east)+(2.8,0.7)$) 
-- ($(chC.north)+(2.7,0.7)$) 
to node[en, left] {$x=10$} (noch);

\end{tikzpicture}
     \caption{Full model of the RAR as used in the feasibility study.}
     \label{fig:automaton_appendix}
\end{figure*}
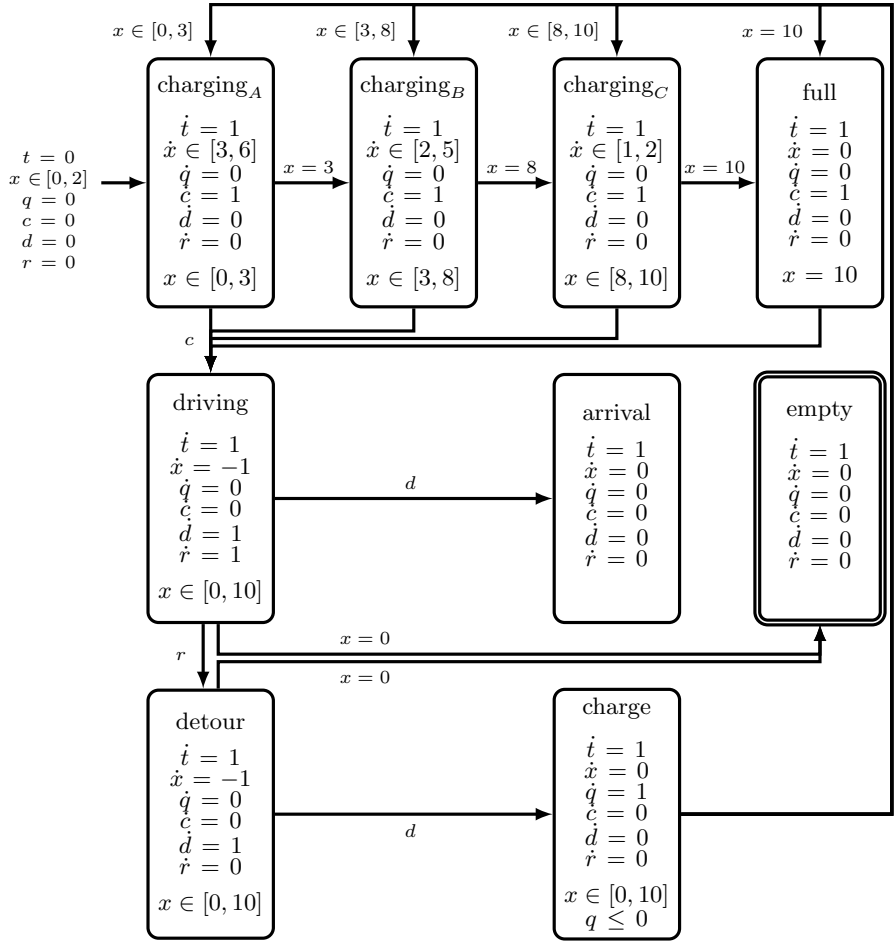

For completeness, we provide the full automaton used for the feasibility study with all variables and flows in Figure \ref{fig:automaton_appendix}. %

\subsection{Validation with \prohver}\label{appendix:prohver}

To the best of our knowledge \prohver is the only tool with the potential to compute a safe overapproximation of prophetic maximum reachability probabilities in these kinds of models.
However, its precision highly depends on the chosen number of discretization intervals, as well as the number of nondeterministic choices in the continuous domain.
Both limit scalability severely, as witnessed by the results computed by \prohver. 
Interval refinement within \prohver requires  integer parameters $(X,Y,Z)$, which bound the discretization and yield the number of discretization intervals (c.f. Table~\ref{appendix:tab:results}).  %
We chose (${60,8,2}$) for $0$ detours and (${10,2,1}$) for $1$ detour. Model complexity grows with the number of detours. For $2$ detours, we could only obtain results for parameters (${5,2,1}$). %
For $0$ detours this resulted in  $66$ intervals and a substantial overapproximation of the reachability probability obtained by \realyst. 
Computation times of \prohver are  between $28$ and $140$ times larger.   
For $1$ detour, \prohver computed $10$ intervals, taking up to \SI{8}{\hour} and resulted in a better approximation of the reachability probabilities. 
For $2$ detours, \prohver was not able to refine the $5$ inital intervals, yielding quick computation times with a substantial overapproximation.
For parameters (${6,2,1}$), \prohver did not terminate within \SI{15}{\hour}.

 \begin{table*}[h]
    \centering
    \scriptsize
    \caption{
      Max reachability probabilities  for  $\Goal=\{(\text{empty}, \val) \mid \val_x =0\}$, $\tmax = (\#\text{detours}+1)\cdot 21$ and $\#\text{detours} \in \{0,1,2\}$.
   Computation times $\comptime$ provided for \realyst (\realystshort) with error $\estat$ and \prohver (\prohvershort) with $\compintervals$ intervals for discretization. 
    $K=(|\VarRandom|, |\reachtree|, |I_S|)$. %
      }
    \label{appendix:tab:results}
    \newcolumntype{Y}{>{\centering\arraybackslash}X}
    \renewcommand{\arraystretch}{1}
    \begin{tabularx}{\linewidth}{lp{0.3cm}p{0.55cm}YYYYYYY}
        \toprule
        \multicolumn{2}{c}{\ }   & \   &    \multicolumn{3}{c}{$0$ detours} &  \multicolumn{3}{c}{$1$ detour}  & $2$ detours  \\
      \cmidrule(lr){4-6}\cmidrule(lr){7-9}\cmidrule(lr){10-10}
            \multicolumn{2}{c}{\ }     & var.    & \texttt{A}& \texttt{AB}& \texttt{ABC}& \texttt{A} &\texttt{AB}& \texttt{ABC}&  \texttt{A}   \\
      \multicolumn{2}{c}{\ }   & $K$ & $(2,8,2)$ &  $(2,12,3)$ & $(2,16,4)$  & $(5,38,10)$ & $(5,88,19)$  & $(5,167,39)$    &  $(8,128,34)$   \\
          \midrule
          \multirow{6}{*}{\rotatebox{90}{rectangular}} 
            & \multirow{3}{*}{\shortstack[l]{\realystshort}} %
            & $p_\textsl{max}$  
            & \probab{0.0503536} & \probab{0.0648239} & \probab{0.0666976} & \probab{0.407864} & \probab{0.42621} & \probab{0.429555} & \probab{0.402105} \\ 
            & & $\estat$ 
            & \errortable{8.76307e-05} & \errortable{0.000119219} & \errortable{0.000128606} & \errortable{0.00171464} & \errortable{0.00167712} & \errortable{0.000613131} & \errortable{0.00332672} \\
            & & {$\comptime$}  
            & \rt{4.49372} & \rt{10.745} & \rt{15.3086} & \rt{484.5} & \rt{1166.49} & \rt{1800.63} & \rt{35570.3} \\
            \cmidrule{2-10} 
            & \multirow{3}{*}{\shortstack[l]{\prohvershort}} %
            & $p_\textsl{max}$  
            & \probab{0.3055917831257279} & \probab{0.32760017561658367} & \probab{0.32760017561658367} & \probab{0.5799221576213734} & \probab{0.6145757767099783} & \probab{0.590174399844356} & \probab{0.7949767298910101} \\
            & &  $\compintervals$
            & \intervals{66} & \intervals{66} & \intervals{66} & \intervals{10} & \intervals{10} & \intervals{10} & \intervals{5} \\
            & & {$\comptime$}  
            & \rt{613.475} & \rt{1202.655} & \rt{1735.179} & \rt{7890.288} & \rt{15247.254} & \rt{28533.796} & \rt{1881.607} \\
          \midrule
          \multirow{6}{*}{\rotatebox{90}{singular}} 
            & \multirow{3}{*}{\shortstack[l]{\realystshort}} %
            & $p_\textsl{max}$  
            & \probab{0.0266728} & \probab{0.0265343} & \probab{0.0265998} & \probab{0.406624} & \probab{0.424759} & \probab{0.429528} & \probab{0.408268} \\
            & & $\estat$
            & \errortable{3.40482e-05} & \errortable{9.47598e-05} & \errortable{0.000112267} & \errortable{0.000711971} & \errortable{0.000670252} & \errortable{0.00167624} & \errortable{0.0037186} \\
            & & {$\comptime$}  
            & \rt{2.46578} & \rt{11.6589} & \rt{19.3762} & \rt{245.213} & \rt{518.298} & \rt{1720.16} & \rt{12187.8} \\
            \cmidrule{2-10} 
            & \multirow{3}{*}{\shortstack[l]{\prohvershort}} %
            & $p_\textsl{max}$  
            & \probab{0.3055917831257279} & \probab{0.32760017561658367} & \probab{0.32760017561658367} & \probab{0.44295903375961115} & \probab{0.4679160428204017} & \probab{0.5239576415909519} & \probab{0.7580344451394565} \\
            & & $\compintervals$
            & \intervals{66} & \intervals{66} & \intervals{66} & \intervals{10} & \intervals{10} & \intervals{10} & \intervals{5} \\
            & & {$\comptime$}  
            & \rt{348.091} & \rt{500.185} & \rt{546.559} & \rt{6984.118} & \rt{10249.278} & \rt{14273.85} & \rt{1461.853} \\
          \bottomrule
          \hspace{0.1cm}
    \end{tabularx}
\end{table*}

\subsection{Computation time distribution}\label{appendix:comptimes}

\definecolor{solBase01}{RGB}{88 110 117}
\definecolor{solMagenta}{RGB}{211  54 130}
\definecolor{color2}{named}{solMagenta}
\definecolor{solGreen}{RGB}{133 153   0}
\definecolor{color3}{named}{solGreen}
\definecolor{solYellow}{RGB}{181 137   0}
\definecolor{color4}{named}{solYellow}
\definecolor{dartmouthgreen}{rgb}{0.05, 0.5, 0.06}
\definecolor{color5}{named}{dartmouthgreen}
\definecolor{solBlue}{RGB}{38 139 210}
\definecolor{color6}{named}{solBlue}
\definecolor{davysgrey}{rgb}{0.33, 0.33, 0.33}
\definecolor{fancyblue}{rgb}{0.01, 0.28, 1.0}

\definecolor{mainColor}{named}{solBase01}
\definecolor{boxcolor}{named}{color4}
\definecolor{cinitloc}{named}{mainColor}
\definecolor{cpath1light}{named}{color6}
\definecolor{cpath1dark}{named}{fancyblue}
\definecolor{cpath2light}{named}{color3}
\definecolor{cpath2dark}{named}{color5}

\tikzstyle{initloc}=[cinitloc]
\tikzstyle{path1start}=[cpath1light,densely dotted]
\tikzstyle{path1end}=[cpath1dark]
\tikzstyle{path2start}=[cpath2light,dotted]
\tikzstyle{path2end}=[cpath2dark]

\definecolor{color-time-forward}{named}{cpath1light}
\definecolor{color-time-processing}{named}{cpath2dark}
\definecolor{color-time-refinement}{named}{color2}
\definecolor{color-time-integration}{named}{amethyst}

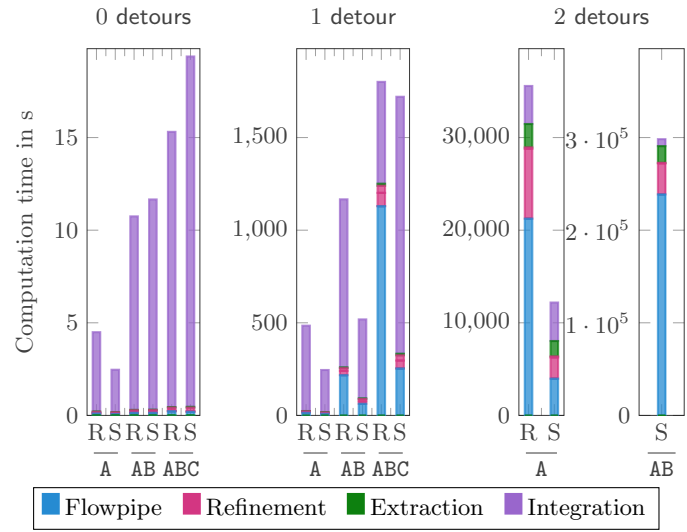
\begin{figure}[h]
    \centering
    \pgfplotstableread[col sep=comma]{plots/data/100kSamples/results_detours0.dat}\datatable

\begin{tikzpicture}[font=\sffamily, fill opacity = 0.75, opacity = 0.75]
    \begin{axis}[colormap/viridis,
    scale only axis=true,
    title=$0$ detours,
        ybar stacked,
        ylabel={Computation time in s}, 
        xticklabels={R,S,R,S,R,S},
xtick=data,
minor x tick num=1,
        enlarge x limits=0.1,
        legend pos=north west,
        grid style=dashed,
        bar width=0.1cm,
        ymin=0,        
        ymax=19.8, %
        width=1.5cm,
        height=0.4\textwidth,
        legend style={
      font=\footnotesize,
      cells={anchor=west},
      legend columns=2,
      at={(0.3,-0.20)},
      anchor=north,
      /tikz/every even column/.append style={column sep=0.2cm}
    },
     draw group line={chargingtype}{A}{\texttt{A}}{-0.5cm}{1pt},
     draw group line={chargingtype}{AB}{\texttt{AB}}{-0.5cm}{1pt},
     draw group line={chargingtype}{ABC}{\texttt{ABC}}{-0.5cm}{1pt}
    ]

    \addplot[color=color-time-processing,  fill, thick, name path=A] table[x=table_index, y=runtime_1, col sep=comma] \datatable;
    \addplot[color=color-time-forward, fill, thick, name path=A] table[x=table_index, y=runtime_21, col sep=comma] \datatable;
    \addplot[color=color-time-refinement,fill,  thick, name path=A] table[x=table_index, y=runtime_22, col sep=comma] \datatable;
    \addplot[color=color-time-refinement, fill, thick, name path=A] table[x=table_index, y=runtime_23, col sep=comma] \datatable;
    \addplot[color=color-time-refinement, fill, thick, name path=A] table[x=table_index, y=runtime_24, col sep=comma] \datatable;
    \addplot[color=color-time-refinement, fill, thick, name path=A] table[x=table_index, y=runtime_25, col sep=comma] \datatable;
    \addplot[color=color-time-refinement,fill,  thick, name path=A] table[x=table_index, y=runtime_26, col sep=comma] \datatable;
    \addplot[color=color-time-processing,fill,  thick, name path=A] table[x=table_index, y=runtime_27, col sep=comma] \datatable;
    \addplot[color=color-time-integration, fill, thick, name path=A] table[x=table_index, y=runtime_28, col sep=comma] \datatable;
    \end{axis}

\end{tikzpicture}
    \pgfplotstableread[col sep=comma]{plots/data/2MSamples/results_detours1.dat}\datatable
\begin{tikzpicture}[font=\sffamily, fill opacity = 0.75, opacity = 0.75]
    \begin{axis}[colormap/viridis,
    scale only axis=true,
    title=$1$ detour,
        ybar stacked,
        xticklabels={R,S,R,S,R,S},
xtick=data,
minor x tick num=1,
        enlarge x limits=0.1,
        legend pos=north west,
        grid style=dashed,
        bar width=0.1cm,
ymin=0,        
ymax=1980, %
        width=1.5cm,
        height=0.4\textwidth,
        legend style={
      font=\footnotesize,
      cells={anchor=west},
      legend columns=2,
      at={(0.3,-0.20)},
      anchor=north,
      /tikz/every even column/.append style={column sep=0.2cm}
    },
     draw group line={chargingtype}{A}{\texttt{A}}{-0.5cm}{1pt},
     draw group line={chargingtype}{AB}{\texttt{AB}}{-0.5cm}{1pt},
     draw group line={chargingtype}{ABC}{\texttt{ABC}}{-0.5cm}{1pt}
    ]

   \addplot[color=color-time-processing,  fill, thick, name path=A] table[x=table_index, y=runtime_1, col sep=comma] \datatable;
     \addplot[color=color-time-forward, fill, thick, name path=A] table[x=table_index, y=runtime_21, col sep=comma] \datatable;
     \addplot[color=color-time-refinement,fill,  thick, name path=A] table[x=table_index, y=runtime_22, col sep=comma] \datatable;
      \addplot[color=color-time-refinement, fill, thick, name path=A] table[x=table_index, y=runtime_23, col sep=comma] \datatable;
      \addplot[color=color-time-refinement, fill, thick, name path=A] table[x=table_index, y=runtime_24, col sep=comma] \datatable;
      \addplot[color=color-time-refinement, fill, thick, name path=A] table[x=table_index, y=runtime_25, col sep=comma] \datatable;
     \addplot[color=color-time-refinement,fill,  thick, name path=A] table[x=table_index, y=runtime_26, col sep=comma] \datatable;
     \addplot[color=color-time-processing,fill,  thick, name path=A] table[x=table_index, y=runtime_27, col sep=comma] \datatable;
     \addplot[color=color-time-integration, fill, thick, name path=A] table[x=table_index, y=runtime_28, col sep=comma] \datatable;
    \end{axis}

\end{tikzpicture}
    \pgfplotstableread[col sep=comma]{plots/data/results_detours2_A.dat}\datatable
\begin{tikzpicture}[font=\sffamily, fill opacity = 0.75, opacity = 0.75]
    \begin{axis}[colormap/viridis,
    name=Ax1,
    scale only axis=true,
        title=$2$ detours,
        title style={xshift=0.8cm},
        ybar stacked,
        xticklabels={R,S},%
        xtick=data,
        minor x tick num=1,
        scaled ticks =false,
        enlarge x limits=0.375,
        legend pos=north west,
        grid style=dashed,
        bar width=0.1cm,
        ymin=0,        
        ymax=39600, 
        width=6mm,
        height=0.4\textwidth,
        legend style={
            font=\footnotesize,
            cells={anchor=west},
            legend columns=4,
            at={(0.3,-0.20)},
            anchor=north,
            /tikz/every even column/.append style={column sep=0.2cm}
        },
        legend to name=named,
         draw group line={chargingtype}{A}{\texttt{A}}{-0.5cm}{1pt},
    ]
      
    \addplot[color=color-time-processing,  fill, thick, name path=A] table[x=table_index, y=runtime_1, col sep=comma,forget plot] \datatable;
    \addplot[color=color-time-forward, fill, thick, name path=A] table[x=table_index, y=runtime_21, col sep=comma] \datatable;
    \addplot[color=color-time-refinement,fill,  thick, name path=A] table[x=table_index, y=runtime_22, col sep=comma,forget plot] \datatable;
    \addplot[color=color-time-refinement, fill, thick, name path=A] table[x=table_index, y=runtime_23, col sep=comma,forget plot] \datatable;
    \addplot[color=color-time-refinement, fill, thick, name path=A] table[x=table_index, y=runtime_24, col sep=comma,forget plot] \datatable;
    \addplot[color=color-time-refinement, fill, thick, name path=A] table[x=table_index, y=runtime_25, col sep=comma,forget plot] \datatable;
    \addplot[color=color-time-refinement,fill,  thick, name path=A] table[x=table_index, y=runtime_26, col sep=comma] \datatable;
    \addplot[color=color-time-processing,fill,  thick, name path=A] table[x=table_index, y=runtime_27, col sep=comma] \datatable;
    \addplot[color=color-time-integration, fill, thick, name path=A] table[x=table_index, y=runtime_28, col sep=comma] \datatable;
    
    \end{axis}
\pgfplotstableread[col sep=comma]{plots/data/10MSamples/results_detours2_AB.dat}\datatable
    \begin{axis}[colormap/viridis,
    at={($(Ax1.north east)+(1cm,0)$)},anchor=north west,
    scale only axis=true,
        ybar stacked,
        xticklabels={S},%
        xtick=data,
        minor x tick num=1,
        scaled ticks =false,
        enlarge x limits=0.75,
        legend pos=north west,
        grid style=dashed,
        bar width=0.1cm,
        ymin=0,        
        ymax=396000, 
        width=6mm,
        height=0.4\textwidth,
        legend style={
            font=\footnotesize,
            cells={anchor=west},
            legend columns=4,
            at={(0.3,0)},
            anchor=north west,
            /tikz/every even column/.append style={column sep=0.2cm}
        },
        legend to name=named,
         draw group line={chargingtype}{AB}{\texttt{AB}}{-0.5cm}{1pt},
    ]
      
    \addplot[color=color-time-processing,  fill, thick, name path=A] table[x=table_index, y=runtime_1, col sep=comma,forget plot] \datatable;
    \addplot[color=color-time-forward, fill, thick, name path=A] table[x=table_index, y=runtime_21, col sep=comma] \datatable;
    \addplot[color=color-time-refinement,fill,  thick, name path=A] table[x=table_index, y=runtime_22, col sep=comma,forget plot] \datatable;
    \addplot[color=color-time-refinement, fill, thick, name path=A] table[x=table_index, y=runtime_23, col sep=comma,forget plot] \datatable;
    \addplot[color=color-time-refinement, fill, thick, name path=A] table[x=table_index, y=runtime_24, col sep=comma,forget plot] \datatable;
    \addplot[color=color-time-refinement, fill, thick, name path=A] table[x=table_index, y=runtime_25, col sep=comma,forget plot] \datatable;
    \addplot[color=color-time-refinement,fill,  thick, name path=A] table[x=table_index, y=runtime_26, col sep=comma] \datatable;
    \addplot[color=color-time-processing,fill,  thick, name path=A] table[x=table_index, y=runtime_27, col sep=comma] \datatable;
    \addplot[color=color-time-integration, fill, thick, name path=A] table[x=table_index, y=runtime_28, col sep=comma] \datatable;
    \legend{Flowpipe, Refinement,Extraction, Integration}
    
    \end{axis}
\end{tikzpicture}
    \\
\ref{named}%

    \caption{Computation times for different models and number of detours with \realyst. We distinguish \emph{flowpipe} (\Cref{subsec:forward}), \emph{refinement} (\Cref{subsec:refinement}), \emph{extraction} (\Cref{subsec:collection}), and \emph{integration} (\Cref{subsec:schedulers}). }
    \label{fig:new_plots}
\end{figure}

As illustrated in Figure~\ref{fig:new_plots} for different model versions, computation time is partitioned into flowpipe construction, refinement operations, extraction of the sample domain and integration. 

The computation time for $0$ detours is dominated by integration, while for more detours the flowpipe construction is predominant. 
On the one hand, more integration samples are required in higher dimensions, since the resulting multidimensional polytope covers a smaller part of the overall integration domain. 
Furthermore, in higher dimensions, it takes longer to evaluate individual samples (containment test). 

On the other hand, with more dimensions all geometrical operations on convex polytopes require a considerably higher computation time, which strongly affects flowpipe construction.
The difference in computation time between the rectangular and singular model variant of the same model version is rooted in the more complex flowpipe construction for rectangular hybrid automata.
This is also illustrated by the fact that for $2$ detours the rectangular variant for model version \texttt{AB} terminated with a memory outage in the flowpipe construction.

Observe that integration time for a fixed number of detours and samples is not necessarily smaller in the singular version of the models, as it exhibits statistical variations due to the underlying importance sampling-based approach.
Model versions \texttt{ABC}, \texttt{AB} and \texttt{A} differ slightly in size, however, not in dimensionality for a given number of detours.
Hence, the impact on computation times is less than the impact of dimensionality.

 \end{appendix}
\label{sec:appendix}

\end{document}